\documentclass[aps,amssymb,groupedaddress,preprint,showpacs,graphics]{revtex4}

\newcommand\ket[1]{\mid#1\rangle}
\newcommand\bra[1]{\langle#1\mid}

\usepackage{graphicx}
\usepackage{dcolumn}
\usepackage{amsmath}

\usepackage{subeqnarray}
\usepackage{bm}
\begin{document}

\title{Nonequilibrium Langevin Approach to\\ Quantum Optics in Semiconductor Microcavities}
\author{S. Portolan$^{1,3,}$\footnote{Electronic address: stefano.portolan@epfl.ch}, O. Di Stefano$^2$, S. Savasta$^2$, F. Rossi$^3$, and R. Girlanda$^2$}
\affiliation{$^1$Institute of Theoretical Physics, Ecole
Polytechnique F\'{e}d\'{e}rale de Lausanne EPFL, CH-1015 Lausanne,
Switzerland} \affiliation{$^2$Dipartimento di Fisica della Materia e
Tecnologie Fisiche Avanzate, Universit\`{a} di Messina Salita
Sperone 31, I-98166 Messina, Italy} \affiliation{$^3$Dipartimento di
Fisica, Politecnico di Torino, Corso Duca degli Abruzzi 24, I-10129
Torino, Italy}

\begin{abstract}
{Recently the possibility of generating nonclassical polariton
states  by means of parametric scattering has been demonstrated.
Excitonic polaritons  propagate in a complex interacting environment
and contain real electronic excitations subject to scattering events
and noise affecting quantum coherence and entanglement. Here we
present a general theoretical framework for the realistic
investigation of polariton quantum correlations in the presence of
coherent and incoherent interaction processes. The proposed
theoretical approach is based on the {\em nonequilibrium quantum
Langevin approach for open systems} applied to interacting-electron
complexes described within the dynamics controlled truncation
scheme. It provides an easy recipe to calculate multi-time
correlation functions which are key-quantities in quantum optics. As
a first application, we analyze the build-up of polariton parametric
emission in semiconductor microcavities including the influence of
noise originating from phonon induced scattering.}

\end{abstract}

\maketitle

\newpage
\section{Introduction}
Entanglement is one of the key features of quantum information and
communication technology\cite{Nielsen-Chuang}. Parametric
down-conversion is the most frequently used method to generate
highly entangled pairs of photons for quantum-optics applications,
such as quantum cryptography and quantum teleportation. Rapid
development in the field of quantum information requires monolithic,
compact sources of nonclassical photon states enabling efficient
coupling into optical fibres and possibly electrical injection.
Semiconductor-based sources of entangled photons would therefore be
advantageous for practical quantum technologies. Moreover
semiconductors can be structured on a nanometer scale, and thus one
may produce materials with tailored properties realizing a wide
variety of physically distinct situations. However semiconductor
heterostructures constitute a complex interacting environment
involving charge, spin, and lattice degrees of freedom, hence suited
to serve as prototype systems where quantum-mechanical properties of
many interacting particles far away from equilibrium can be studied
in a controlled fashion\cite{AxtKuhn}. It has been demonstrated that
very large $\chi^{(3)}$ resonant polaritonic nonlinearities in
wide-gap semiconductors and in semiconductor microcavities can be
used to achieve parametric emission\cite{CuCl Honerlage,Baumberg}.

Polaritons are mixed quasiparticles resulting from the strongly
coupled propagation of light and collective electronic excitations
(excitons) in semiconductor crystals. Although spontaneous
parametric processes involving polaritons in bulk semiconductors
have been known for decades\cite{CuCl Honerlage}, the possibility of
generating entangled photons by these processes was theoretically
pointed out only lately\cite{Savasta PRL96}. This result was based
on a microscopic quantum theory of the nonlinear optical response of
interacting electron systems relying on the dynamics controlled
truncation scheme\cite{Victor-Axt-Stahl PRB} extended to include
light quantization\cite{Savasta PRL2003,HRS Savasta,SSC Savasta}.
The above theoretical framework was also applied to the analysis of
polariton parametric emission in semiconductor microcavities
(SMCs)\cite{Savasta PRL2003,HRS Savasta}. A SMC is a photonic
structure designed to enhance light-matter interactions. The strong
light-matter interaction in these systems gives rise to cavity
polaritons which are hybrid quasiparticles consisting of a
superposition of cavity photons and quantum well excitons
\cite{Weisbuch-Houdre}. Demonstrations of parametric amplification
and parametric emission in SMCs\cite{Baumberg, Erland, Langbein
PRB2004}, together with the possibility of ultrafast optical
manipulation and ease of integration of these microdevices, have
increased the interest on the possible realization of nonclassical
cavity-polariton states\cite{squeezing Quattropani,CiutiBE,Savasta
PRL2005,LosannaCC,SSC Savasta}. In 2004, experimental evidence for
the generation of ultraviolet polarization-entangled photon pairs by
means of biexciton resonant parametric emission in a single crystal
of semiconductor CuCl has been reported\cite{Nature CuCl}.
Short-wavelength entangled photons are desirable for a number of
applications as generation of further entanglement between three or
four photons. In 2005 an experiment probing quantum correlations of
(parametrically emitted) cavity polaritons by exploiting quantum
complementarity has been proposed and realized\cite{Savasta
PRL2005}. Specifically, it has been shown that polaritons in two
distinct idler modes interfere if and only if they share the same
signal mode so that which-way information cannot be gathered,
according to Bohr's quantum complementarity principle. In 2006 a
promising low-threshold parametric oscillation in vertical triple
SMCs with signal, pump and idler waves propagating along the
vertical direction of the nanostructure has been
demonstrated\cite{Ciuti Nature}.

The crucial role of many-particle Coulomb correlations in
semiconductors marks a profound difference from dilute atomic
systems, where the optical response is well described by independent
transitions between atomic levels, and the nonlinear dynamics is
governed only by saturation effects due to the Pauli exclusion
principle. In planar SMCs, thanks to their mutual Coulomb
interaction, pump polaritons generated by resonant optical pumping
may scatter into pairs of polaritons (signal and idler)\cite{Savasta
PRL96,Baumberg,Ciuti parlum}, they are determined by the two
customary energy and wave vector conservation conditions $2{\bf k}_p
= {\bf k}_s + {\bf k}_i$ and $2 E_{{\bf k}_p} = E_{{\bf k}_s} +
E_{{\bf k}_i}$ depicting an eight-shaped curve in momentum space. At
low pump intensities they are expected to undergo a spontaneous
parametric process driven by vacuum-fluctuation, whereas at moderate
intensities they display self-stimulation and oscillation
\cite{Baumberg}. However they are real electronic excitations
propagating in a complex interacting environment. Owing to the
relevance of polariton interactions, and also owing to their
interest for exploring quantum optical phenomena in such a complex
environment, theoretical approaches able to model accurately
polariton dynamics including light quantization, losses and
environment interactions are highly desired. The analysis of
nonclassical correlations in semiconductors constitutes a
challenging problem, where the physics of interacting electrons must
be added to quantum optics and should include properly the effects
of energy relaxation, dephasing, and noise, induced by
electron-phonon interaction \cite{Kuhn-Rossi PRB 2005}.

Previous descriptions of polariton parametric processes make deeply
use of the picture of polaritons as interacting bosons. These
theories have been used to investigate parametric amplifications,
parametric luminescence, coherent control, entanglement and
parametric scattering in momentum space\cite{Ciuti parAmpl,Ciuti
parlum, LosannaCC,CiutiBE,Langbein PRB2004}.

It is worth noting that in a realistic environment phase-coherent
nonlinear optical processes involving real excitations compete with
incoherent scattering as evidenced by experimental results. In
experiments dealing with parametric emission, what really dominates
emission at low pump intensities is the photoluminescence (PL) due
to the incoherent dynamics of the single scattering events driven by
the pump itself and  the Rayleigh scattering of the pump due to the
unavoidable presence of structural disorder. The latter process is
elastic and can thus be spectrally filtered in principle, moreover
it is confined in k-space to a ring of in-plane wave vectors with
almost the same modulus of the pump wave vector. On the contrary PL,
being not an elastic process, cannot be easily separated from
parametric emission. Only once the pumping become sufficient the
parametric processes start to reveal themselves and to take over
pump-induced PL as well. Indeed, usually, parametric emission and
standard pump-induced PL cohabit as shown by experiments at low and
intermediate  excitation density\cite{Langbein PRB2004}. Moreover,
in order to address quantum coherence properties and
entanglement\cite{Nature CuCl} the preferred experimental situations
are those of few-particle regimes, namely coincidence detection in
photon counting. In this regime, the presence of incoherent noise
due to pump-induced PL tends to spoil the system of its coherence
properties lowering the degree of nonclassical correlations. The
detrimental influence of incoherent effects on the quantum coherence
properties is also well evidenced in the measured time-resolved
visibility shown in Ref. \onlinecite{Savasta PRL2005}. At initial
times visibility is suppressed until  parametric emission prevails.
Thus, a microscopic analysis able to account for parametric emission
and pump-induced PL on an equal footing is highly desirable in order
to make quantitative comparison with measurements and propose future
experiments. Furthermore a quantitative theory would be of paramount
importance for a deeper understanding of quantum correlations in
such structures aiming at seeking and limiting all unwanted
detrimental contributions.

The dynamics controlled truncation scheme (DCTS) provides a (widely
adopted) starting point for the microscopic theory of light-matter
interaction effects beyond mean-field \cite{AxtKuhn,Victor-Axt-Stahl
PRB}, supplying a consistent and precise way to stop the infinite
hierarchy of higher-order correlations which always appears in the
microscopic approaches of manybody interacting systems. In 1996 the
DCTS was extended in order to include in the description the
quantization of the electromagnetic field \cite{Savasta PRL96}. This
extension has been applied to the study of quantum optical phenomena
in semiconductors as polariton entanglement \cite{SSC Savasta}.
However, in these works damping has been considered only at a
phenomenological level.

In this paper we shall present a novel approach based on a
DCTS-nonequilibrium quantum Langevin description of the open system
in interaction with its surroundings. This approach enables us to
include on an equal footing the microscopic description of the
scattering channels competing with the coherent parametric phenomena
the optical pump induces. We shall apply our method in order to
perform a more realistic description of light emission taking into
account nonlinear parametric interactions, light quantization,
cavity losses and polariton-phonon interaction. The developed
theoretical framework can be naturally extended to include other
incoherent scattering mechanisms such as the interaction of
polaritons with thermal free electrons \cite{Di Carlo Kavokin PRB}.
As a first application of the proposed theoretical scheme, we have
analyzed the time-resolved and  time-integrated build-up of
polariton parametric emission in semiconductor microcavities
including the influence of noise originating from phonon induced
scattering. The presented numerical results clearly evidence the
role of incoherent scattering in parametric photoluminescence and
thus show the importance of a proper microscopic analysis able to
account for parametric emission and pump-induced PL on an equal
footing. We also exploit the present approach to calculate the
emission spectra as a function of the pump power density. The
spectra  display a significant line-narrowing as well as parametric
emission starts to prevail.

The paper is organized as follows. In Sec \ref{DCTS phenom},
starting from a DCTS theory for semiconductor microcavities
\cite{nostro teoria}, we present a theory of $\chi^{(3)}$ optical
nonlinearities in terms of interacting polaritons. The latter
focuses mainly on the nonlinear part in order to model coherent
optical parametric  processes and the damping is included only
phenomenologically. In Section \ref{III} we apply a nonequilibrium
quantum Langevin treatment of damping and fluctuations in an open
system, originally proposed by Lax. Section \ref{IV} will be devoted
to the microscopic calculation of phonon-induced scattering rates
and polariton PL within a second order Born-Markov approximation. In
Sec. \ref{V} we shall present a quantum Langevin description of
parametric emission including incoherent effects; particular
attention will be devoted to the case of single pump feed, whose
results will be the subject of Sec. \ref{VI}. Finally in Sec.
\ref{VII} we shall summarize and draw some conclusions.
\section{Dynamics Controlled Truncation Scheme for Interacting
Polaritons}\label{DCTS phenom}

The system under investigation consists of one or more (uncoupled)
QWs grown inside a semiconductor planar Fabry-Perot resonator. For
the quasi-2D interacting electron system we adopt the usual
semiconductor model Hamiltonian \cite{AxtKuhn} which can be
expressed as
\begin{equation}\label{Ham electron} \hat{H}_e =  \sum_{N \alpha}
\hbar \omega_{N \alpha} \ket{{N \alpha}} \bra{{N \alpha}}\, ,
\end{equation} where the eigenstates of $\hat{H}_e$ have been
labeled according to the number N of electron-hole ({\em eh}) pairs.
The state $\ket{{N=0}}$ is the electronic ground state, the $N=1$
subspace is the exciton subspace with the additional collective
quantum number $\alpha$ denoting the exciton energy level $n$ and
the in-plane wave vector ${\bf k}$. The set of states with $N = 2$
determines the biexciton subspace. We treat the planar-cavity field
within the quasimode approximation, the cavity field is quantized as
though the mirrors were perfect:
\begin{equation}\label{Ham cavity} \hat{H}_c =
\sum_k \hbar \omega_{k} \hat{a}_{\bf k} ^{\dag} \hat{a}_{\bf k}\, ,
\end{equation} and the resulting discrete modes are then coupled to
the external continuum of modes by an effective Hamiltonian
\begin{equation}\label{H quasi modi} \hat{H}_p = i \hbar \, t_{c,j} \sum_{j=1,2, \bf k} (\hat{E}^{(-)}_{j,\bf k} \hat{a}^\dag_{\bf k} -
\hat{E}^{(+)}_{j,\bf k} \hat{a}_{\bf k})\, ,\end{equation} where $j$
labels the two mirrors and $t_j$ determines the fraction of the
field amplitude passing the cavity mirror, $\hat{E}^{(-)}_{j,\bf k}$
($\hat{E}^{(+)}_{j,\bf k}$) is the positive (negative) frequency
part of the coherent input light field. The coupling of the electron
system to the cavity modes is given within the usual rotating wave
approximation
\begin{equation}\label{Ham inter cav-exc} \hat{H}_{I} = - \sum_{n {\bf k}}
\hbar V_{n {\bf k}} \hat{a}_{\bf k} ^{\dag} \hat{B}_{n {\bf k}} +
H.c. \end{equation} $\hat{B}_{n {\bf k}}$ is the exciton destruction
operator and can be expanded as well in terms of the energy
eigenstates of the electron system. For later convenience, the
exciton and photon operators are normalized so that $\hat
B^\dag_{\bf k}\hat B_{\bf k}$ and $\hat a^\dag_{\bf k} \hat a_{\bf
k}$ are operators corresponding to the number of particles within a
Bohr-radius two-dimensional disk ($\pi a^2_\text{x}$) at a given
${\bf k}$.

We start from the Heisenberg equations of motion for the exciton and
photon operators. In the DCTS spirit, we keep only those terms
providing the lowest nonlinear response ($\chi^{(3)}$) in the input
light field \cite{nostro teoria}. We assume the pump polaritons
driven by a quite strong coherent input field ${E}_{\bf
k}^{in}=\langle \hat E^{(-)}_{1{\bf k}} \rangle$ consisting of a
classical ($\mathbb{C}$-number) field, resonantly exciting the
structure at a given energy and wave vector, $\bf{k}_p$. We are
interested in studying polaritonic effects in SMCs where the optical
response involves mainly excitons belonging to the 1S band with wave
vectors close to normal incidence, $|{\bf k}| \ll
\frac{\pi}{a_{\text{x}}}$. We retain only those terms containing the
semiclassical pump amplitude twice, thus focusing on the ``direct"
pump-induced nonlinear parametric interaction. One ends up with a
set of coupled equations of motion exact to the third order in the
exciting field. While a systematic treatment of higher-order optical
nonlinearities would require an extension of the equations of
motion, a restricted class of higher-order effects can be obtained
from solving these equations self-consistently up to arbitrary order
as it is usually employed in standard nonlinear optics. This can be
simply accomplished by replacing, in the nonlinear sources, the
linear excitonic polarization and and light field operators with the
total field. From now on, that the pump-driven terms (e.g. the $B$
and $a$ at ${\bf k}_p$) are ${\mathbb{C}}$-numbers coherent
amplitudes like the semiclassical electromagnetic pump field, we
will make such distinction in marking with a ``hat" the operators
only. It yields \cite{nostro teoria}
\begin{subeqnarray}\label{2}
  \dot{\hat{B}}_{\bf k}&=&
  -i{{\omega}}^\text{x}_{\bf k}  \hat{B}_{\bf k} -i \hat{s}_{\bf k}
  +i V \hat{a}_{\bf k}-i\hat{R}^{NL}_{\bf k}\, ,\\
\dot{\hat{a}}_{\bf k}&=&
  -i{{\omega}}^c_{\bf k}\hat{a}_{\bf k}
  +i V\hat{B}_{\bf k}+ t_c {E}_{\bf k}^{in}\, \,
  ;
\end{subeqnarray}
where ${\omega}^{(i)}_{\bf k}$  ($i= \text{x},c$) are the energies
of  QWs excitons and cavity photons. The intracavity and the exciton
field of a given mode ${\bf k}$ are coupled by the exciton-cavity
photon coupling rate $V$. The relevant non-linear source term, able
to couple waves with different in-plane wave vector ${\bf k}$, is
given by $\hat{ R}^{NL}_{\bf k}=(\hat{ R}^{sat}_{\bf
k}+\hat{R}^{\text{xx}}_{\bf k})/N_{eff}$; where the first term
originates from the phase-space filling of the exciton transition,
\begin{equation}\label{Rsat}
\hat{R}^{sat}_{{\bf k}} = \frac{V}{n_{\text{sat}}} B_{{\bf k}_p} a_{
{\bf k}_p} \hat{B}^\dag_{{\bf k}_i}\, ;
\end{equation}
being ${n_{sat}}= 7/16$ the exciton saturation density and ${\bf
k}_i = 2{\bf k}_p-{\bf k}$. $N_{eff}$ depends on the number of wells
inside the cavity and their spatial overlap with the cavity-mode.
Inserting a large number of QWs into the cavity results also in
increasing the photon-exciton coupling rate $V= V_1 \sqrt{N_{eff}}$,
where $V_1$ is the exciton-photon coupling for 1 QW.
 ${\hat R}^{\text{xx}}_{\bf k}$ is the Coulomb interaction term. It
dominates the coherent xx coupling and for co-circularly polarized
waves (the only case here addressed) can be written as
\begin{eqnarray}
&&\hat{R}^{\text{xx}}_{{\bf k}} = \hat{B}^\dag_{{\bf k}_i}(t) \bigg(
V_{\text{xx}} B_{{\bf k}_p}(t) B_{{\bf k}_p}(t) - \nonumber \\
&& - i \int_{-\infty}^t dt' F^{}(t-t') B_{{\bf k}_p}(t') B_{{\bf
k}_p}(t') \bigg)\, , \label{Rxx}\end{eqnarray} where $V_{\text{xx}}
\simeq 6 E_b / \pi$, being $E_b$ the exciton binding energy.
Equation\ (\ref{Rxx}) includes the instantaneous mean-field xx
interaction term  and  a non-instantaneous term originating from
four-particle correlations. These equations show a close analogy to
those derived in \cite{HRS Savasta}, addressing the bulk case. In
addition to that former result, in the present formulation we
succeed in dividing rigorously (in the DCTS spirit) the
Coulomb-induced correlations into mean-field and four-particle
correlation terms. Moreover the pump-induced shift due to parametric
scattering $\hat{s}_{\bf k}$ reads
\begin{eqnarray}
&&N_{eff} \hat{s}_{\bf k} = \frac{V}{n_{\text{sat}}} \bigg(B^*_{{\bf
k}_p} a_{{\bf k}_p} \hat{B}_{{\bf k}} + B^*_{{\bf k}_p} B_{{\bf
k}_p} \hat{a}_{{\bf k}} \bigg) +2 V_{\text{xx}} B^*_{{\bf k}_p} B_{
{\bf
k}_p} \hat{B}_{{\bf k}}- \nonumber \\
&& \hspace{2.5cm} -2 i B^*_{{\bf k}_p}(t) \int_{-\infty}^t dt'
F^{}(t-t') \hat{B}_{{\bf k}}(t') B_{{\bf k}_p}(t') \,
.\end{eqnarray} Equation (\ref{2}) can be written in compact form as
\begin{equation}\label{dif1}
\dot {\mathbf{\cal B}}_{\bf k} =
-i\mathbf{ \Omega}^{\text{xc}}_{\bf k}\,
\mathbf{\cal B}_{\bf k} + \mathbf{\cal E}^{in}_{\bf k}
-i \mathbf{{\cal R}}_{\bf k}^{NL};
\end{equation}
where $ \mathbf{\cal B}_{\bf k} \equiv \left(
\begin{array}{c}
\hat{B}_{\bf k} \\
\hat{a}_{\bf k}
\end{array} \right)
$,
$
  \mathbf{\Omega}^{\text{xc}}_{\bf k} \equiv \left( \begin{array}{cc}
{\omega}^x_{\bf k} &  -V\\
-V &{\omega}^c_{\bf k}
\end{array} \right)$,
$ \mathbf{\cal E}^{in}_{\bf k} \equiv \left(
\begin{array}{c}
0 \\
t_c {E}^{in}_{\bf k}
\end{array} \right)
$, and $ \mathbf{{\cal R}}_{\bf k}^{NL} \equiv \left(
\begin{array}{c}
\hat{R}^{NL}_{\bf k} \\
0
\end{array} \right)
$.
When the coupling rate $V$ exceeds the  decay rate of the exciton
coherence and of the cavity field, the system enters the strong
coupling regime. In this regime, the continuous exchange of energy
before decay significantly alters the dynamics and hence the
resulting resonances  of the coupled  system  with respect to those
of bare excitons and cavity photons. As a consequence,
cavity-polaritons arise as the two-dimensional eigenstates of
$\mathbf{\Omega}^{\text{xc}}_{\bf k}$. The coupling rate $V$
determines the splitting ($\simeq 2V$) between the two polariton
energy bands. This nonperturbative dynamics including the
interactions (induced by $\hat{R}^{NL}_{\bf k}$) between different
polariton modes can be accurately described by Eq.\, (\ref{2}).
Nevertheless there can be reasons to prefer a  change of bases from
excitons and photons to the eigenstates of the coupled system,
namely polaritons. An interesting one is that the resulting
equations may provide a more intuitive description of nonlinear
optical processes in terms of interacting polaritons. Moreover
equations describing the nonlinear interactions between polaritons
become more similar to those describing parametric interactions
between photons widely adopted in quantum optics. Another, more
fundamental reason, is that the standard second-order Born-Markov
approximation scheme, usually adopted to describe the interaction
with environment, is strongly bases-dependent, and using the
eigenstates of the closed system provides more accurate results. In
order to obtain  the dynamics for the polariton system we perform on
the exciton and photon operators the unitary basis transformation
\begin{equation}\label{base}
\mathbf{\cal P}_{\bf k}=U_{\bf k}\mathbf{\cal B}_{\bf k};
\end{equation}
being $\mathbf{\cal P}_{\bf k}=\left( \begin{array}{c}
\hat{P}_{1{\bf k}} \\
\hat{P}_{2{\bf k}} \end{array} \right)$ and
\begin{equation}\label{base2}
U_{\bf k}=\left( \begin{array}{cc}
X_{1{\bf k}} & C_{1{\bf k}}\\
X_{2{\bf k}} & C_{2{\bf k}} \end{array} \right)\, .
\end{equation}
In general photon operators obey Bose statistics, on the contrary
the excitons do not posses a definite statistics (i.e. either
bosonic or fermionic), but their behaviour may be well approximated
by a bosonic-like statistics in the limit of low excitation
densities. Indeed
\begin{equation}\label{[]}
[\hat{B}_{n }, \hat{B}^\dag_{n'} ] = \delta_{n',n} - \sum_{\bf q}
\Phi^{*}_{n {\bf q}} \Phi^{}_{n' {\bf q}} \sum_{N,\alpha,\beta}
\Biggl(\bra{N \alpha}\hat{c}^\dag_{{\bf q}} c_{{\bf q}}\ket{N \beta}
+ \bra{N \alpha}\hat{d}^\dag_{-{\bf q}} d_{-{\bf q}}\ket{N \beta}
\Biggr)  \ket{N \alpha} \bra{N \beta}\, .
\end{equation}
Thus, within a DCTS line of reasoning \cite{Sham PRL95}, the
expectation values of these transition operators (i.e. $\ket{N
\alpha} \bra{N \beta}$) are at least of the second order in the
incident light field, they are density-dependent contributions.
Evidently all these consideration affect polariton statistics as
well, being polariton linear combination of intracavity photons and
excitons. As a consequence, even if polariton operators have no
definite statistics, in the limit of low excitation intensites they
obey approximately bosonic-like commutation rules.

Diagonalizing $\mathbf{\Omega}^{\text{xc}}_{\bf k}$:
\begin{equation}\label{diagonalization}
U_{\bf k} \mathbf{\Omega}^{\text{xc}}_{\bf k} = \tilde \Omega_{\bf
k} U_{\bf k}\, ,
\end{equation}
where
\[
\tilde \Omega_{\bf k}=\left( \begin{array}{cc}
{\omega}_{1{\bf k}} & 0\\
0 & {\omega}_{2{\bf k}} \end{array} \right).
\]
$\omega_{1,2}$ are the eigenenergy (as a function of ${\bf k})$ of
the lower (1) and upper (2) polariton states. After simple algebra
it is possible to obtain this relation for the Hopfield coefficients
\cite{Ciuti SST}:
\begin{equation}
X_{1{\bf k}}=-C_{2{\bf k}}^*;\qquad  C_{1{\bf k}}=X_{2{\bf k}}^*\, .
\end{equation}
where
\begin{equation}
X_{1{\bf k}}= \frac{1}{\sqrt{1+\left(\frac{V}{{\omega}_{1{\bf k}} -
\omega^c_k}\right)^2}}\qquad C_{1{\bf
k}}=\frac{1}{\sqrt{1+\left(\frac{{\omega}_{1{\bf k}} -
\omega^c_k}{V}\right)^2}} \, .
\end{equation}

Introducing this transformation into Eq.\, (\ref {dif1}), one
obtains
\begin{equation}\label{dif2}
\dot {\mathbf{\cal P}_{\bf k}}=-i
\tilde \Omega_{\bf k}
\mathbf{\cal P}_{\bf k} + \mathbf{\tilde {\cal E}}^{in}_{\bf k}
-i \mathbf{\tilde {\cal R}}_{\bf k}^{NL};
\end{equation}
where $\mathbf{\tilde {\cal R}}^{NL}=U \mathbf{ {\cal R}}^{NL}$,
which in explicit form reads
\begin{subeqnarray}\label{3}
  \dot{\hat{P}}_{1\bf k}&=&
  -i{{\omega}}_{1\bf k} {\hat{P}}_{1\bf k} -i {\tilde s_{1
\bf k}} + {\tilde E}^{in}_{1,\bf k}-i {\tilde R}^{NL}_{1\bf k}\, ,\\
\dot{\hat{P}}_{2\bf k}&=&
  -i{{\omega}}_{2\bf k}{\hat{P}}_{2\bf k} -i {\tilde s_{2
\bf k}} +{\tilde E}^{in}_{2,\bf k}-i {\tilde R}^{NL}_{2\bf k}\, \,
  ;
\end{subeqnarray}
where ${\tilde E}^{in}_{m\bf k}=t_cC^{}_{m\bf k}{E}_{\bf k}^{in}$,
and ${\tilde R}^{NL}_{m\bf k}=X_{i\bf k}\hat{R}^{NL}_{\bf k}$,
$(m=1,2)$. Such a diagonalization is the necessary step when the
eigenstates of the polariton system are to be used used as the
starting states perturbed by the interaction with the environment
degrees of freedom \cite{Piermarocchi bottleneck}. The nonlinear
interaction written in terms of polariton operators reads
\begin{equation}\label{nonlinear}
  \hat{R}^{NL}_{\bf k}=\hspace{-0.2 cm} \sum_{i, j,l}{\hat{P}}^\dag_{i {\bf k}_i}(t)
  \int_{-\infty}^{t} \hspace{-0.25 cm} g^{i j l}_{m {\bf k}}(t,t')P_{j {\bf k}_p}(t') P_{l{\bf
  k}_p}(t')dt'\,,
\end{equation}
being
\begin{eqnarray}\label{g}
  && g_{m {\bf k}}^{ijl}(t,t') = \frac{1}{N_{eff}}
\bigg[ \frac{V}{n_{sat}} C^*_{j,{\bf k}_p} \delta(t-t')+  \nonumber \\
&& \bigg( V_{\text{xx}} \delta(t-t') -i F(t-t') \bigg) X^*_{j,{\bf
k}_p} \bigg] X^*_{l,{\bf k}_p} X_{i,{\bf k}_i}\, .
\end{eqnarray}
The shift $\hat{s}_{\bf k}(t)$ is transformed into
\begin{equation} \tilde
s_{m \bf k}(t) = \sum_{i j l} P^*_{i {\bf k}_p}(t)
\int_{-\infty}^{t} \hspace{-0.25 cm} \bigg( h^{i j l}_{m \bf k}
\delta(t-t') - 2i F(t-t') \bigg) P_{j {\bf k}_p} (t') \hat{P}_{l
{\bf k}}(t') dt' \, ,
\end{equation} and
\begin{eqnarray}
h^{i j l}_{m \bf k} = \frac{1}{N_{eff}} X_{m \bf k} \bigg[
\frac{V}{n_{sat}} X_{i{\bf k}_p} \bigg( C^*_{j,{\bf k}_p} X^*_{{l
\bf k}} + X^*_{j{\bf k}_p} C^*_{l{\bf k}} \bigg) + \nonumber \\
+ 2 V_{\text{xx}} X_{i{\bf k}_p} X^*_{j{\bf k}_p} X^*_{l{\bf k}}
\bigg]\, .
\end{eqnarray}

Equation\ (\ref{3}) describes the coherent dynamics of a system of
interacting cavity polaritons. The nonlinear term drives the mixing
between polariton modes with different in-plane wave vectors and
possibly belonging to different branches. Of course there are
nonlinear optical processes involving modes of only one
branch\cite{SSC Savasta,Savasta PRL2005}. In this case it is
possible to take into account only one of the two set of equations
in (\ref{3}) and to eliminate the summation over the branch indexes
in Eq.\, (\ref{nonlinear}).

Equations (\ref{3}) can be considered the starting point for the
microscopic description of quantum optical effects in SMCs. They
extend the usual semiclassical  description of Coulomb interaction
effects, in terms of a mean-field term plus a genuine
non-instantaneous four-particle correlation, to quantum optical
effects. Only the many-body electronic Hamiltonian, the
intracavity-photon Hamiltonian and the Hamiltonian describing their
mutual interaction have been taken into account. The proper
microscopic inclusion of losses through mirrors, decoherence and
noise due to environment interactions will be the main subject of
the following sections.

\section{Quantum Langevin noise sources : Lax theorem}\label{III}
In order to model the quantum dynamics of the polariton system in
the presence of losses and decoherence we  exploit the microscopic
quantum Heisenberg-Langevin approach. We choose it because of its
easiness in manipulating operators differential equations, and above
all, for its invaluable flexibility and strength in performing even
multitime correlation calculations, so important when dealing with
quantum correlation properties of the emitted light. Moreover, as we
shall see in the following, it enables, under certain assumptions, a
(computationally advantageous) decoupling of  incoherent dynamics
from  parametric processes.

In the standard well-known theory of quantum Langevin noise
treatment \cite{Ford,Mandel-Wolf} greatly exploited in quantum
optics, one uses a perturbative description and thanks to a Markov
approximation gathers the damping as well as a term including the
correlation of the system with the environment. The latter arises
from the initial values of the bath operators, which are assumed to
behave as noise sources of stochastic nature. Normally the model
considered has the form of harmonic oscillators coupled linearly to
a bosonic environment. The standard statistical viewpoint is easy
understood: the unknown initial values of the bath operators are
considered as responsible for fluctuations, and the most intuitive
idea is to assume bosonic commutation relations for the Langevin
noise sources because the bath is bosonic too. Most times these
commutation relations are introduced phenomenologically with damping
terms taken from experiments and/or from previous works. In other
contexts a microscopic calculation has been attempted using a
quantum operator approach. Besides its valuable results as soon as
one tries to set a microscopic calculation for interaction forms
different from a 2-body linear coupling \cite{Mandel-Wolf}, e.g.
acoustic-phonon interaction, some problems arise and one is lead to
consider additional approximations in order to close the equations
of motion and obtain damping and fluctuations.

In 1966 Melvin Lax, with clear in mind the lesson of classical
statistical mechanics of Brownian motion, extended the noise-source
technique to quantum systems. In general, the model comprises a
system of interest coupled to a reservoir (R). Considering a generic
global (i.e. system+reservoirs) operator, a first partial trace over
the reservoir degrees of freedom results in still a system operator,
a subsequent trace over the systems degrees of freedom would give an
expectation value. In order to be as clear as possible we shall
denote the former operation on the environment by single brackets
$\langle \ \ \rangle_R$, whereas for the combination of the two
(partial trace over the reservoir and subsequent partial trace over
the system  density matrices) the usual brackets $\langle \ \
\rangle$ is used. His philosophy was that the reservoir can be
{\emph completely eliminated} provided that frequency shift and
dissipation induced by the reservoir interactions are incorporated
into the mean equations of motion, and provided that suitable
operator noise sources with the correct moments are added. In Ref.\,
\cite{Lax} he proposed for the first time that as soon as one is
left with a closed set of equations of system operators for the mean
motion ({\emph mean} with respect to the reservoir) they can be
promoted to equations for global bare operators (system+reservoir)
provided to consider additive noise sources endowed by the proper
statistics due to the system dynamics. He showed that in a Markovian
environment these noise source operators must fulfill generalized
Einstein equations which are a sort of time dependent
non-equilibrium fluctuation-dissipation theorem.

If ${\bf \hat{a}} = \{\hat{a}_1,\hat{a}_2, \cdot \cdot \cdot \}$ is
a set of system operators, and \begin{equation}\label{lax mean eq}
\frac{d \left<\right. \hat{a}_{\mu} \left.\right>_R}{dt} =
\left<\right. \hat{A}_{\mu}({\bf \hat{a}})\left.\right>_R
\end{equation} are the correct equations for the mean, then one can
show that the equations
\begin{equation} \frac{d \hat{a}_{\mu} }{dt} = \hat{A}_{\mu}({\bf \hat{a}}) +
\hat{F}_{\mu}({\bf \hat{a}},t) \end{equation} are a valid set of
equations of motion for the operators provided the additive noise
operators $\hat{F}$'s to be endowed with the correct statistical
properties to be determined for the motion itself.

The Langevin noise source operators are such that their expectation
values $\langle \hat{\mathcal{F}}_{\mu} \rangle_R$ vanish, but their
second order moments do not \cite{Lax}. They are intimately linked
up with the global dissipation and in a Markovian environment they
take the form:
\begin{equation}\label{Fluc-Diss}
\hspace{-1.0cm} \langle \hat{\mathcal{F}}_{\mu} (t)
\hat{\mathcal{F}}_{\nu}(u) \rangle_R = 2 \langle \hat{D}_{\mu \nu}
\rangle_R \, \delta(t-u)\, , \end{equation} where the diffusion
coefficients are
\begin{equation}\label{Diff coeff def} 2\langle \hat{D}_{\mu \nu}
\rangle_R = \frac{d}{d t} \langle \hat{a}_{\mu}(t) \hat{a}_{\nu}(t)
\rangle_R - \langle {\Big \{} \frac{d}{d t} \hat{a}_{\mu} {\Big
\}}\hat{a}_{\nu} \rangle_R - \langle \hat{a}_{\mu} {\Big \{}
\frac{d}{d t} \hat{a}_{\nu} {\Big \}} \rangle_R \, ,
\end{equation}
\begin{equation}
{\Big \{} \frac{d}{d t} \hat{a}_{\nu} {\Big \}} \equiv \frac{d}{d t}
\hat{a}_{\nu} - \hat{\mathcal{F}}_{\nu}\, .
\end{equation}
Equation (\ref{Fluc-Diss}) is an (exact) time dependent Einstein
equation representing a fluctuation-dissipation relation valid for
nonequilibrium situations, it witnesses the fundamental
correspondence between dissipation and noise in an open system.
$\langle \hat{D}_{\mu \nu} \rangle_R$ becomes not only
time-dependent, it is a system operator and can be seen as the
extent to which the usual rules for differentiating a product is
violated in a Markovian system. Equation\ (\ref{Fluc-Diss}) and
Eq.\, (\ref{Diff coeff def}) make the resulting
``fluctuation-dissipation" relations between $\hat{D}_{\mu \nu}$ and
the reservoir contributions to be in precise agreement with those
found by direct use of perturbation theory. This method, however,
guarantees the commutation rules for the corresponding operators to
be necessarily preserved in time. This result is more properly an
exact, quantitative, theorem which gives relevant insights regarding
the intertwined microscopic essence of damping and fluctuations in
any open system.

In order to be more \textit{specific}, let us consider a single
semiclassical pump feed resonantly exciting the lower polariton
branch at a given wave vector ${\bf k}_p$. It is worth noticing,
however, that the generalization to a many-classical-pumps settings
is straightforward. The nonlinear term $R^{NL}$ of Eq.\,
(\ref{nonlinear}) couples pairs of wave vectors, let's say $ {\bf
k}$, the signal, and $ {\bf k}_i= 2{\bf k}_p - {\bf k}$, the idler.
A  general result for quantum systems interacting with a Markovian
environment is that after tracing over the bath degrees of freedom
ones remains with system equations of motion in the bence of the
environment plus additional phase-shifts (often neglected) and
relaxation terms \cite{Lax}. The Heisenberg Eqs.\, (\ref{3}),
involving system operators, for the generic couple read
\begin{eqnarray}\label{sist
lineare val asp} \frac{d}{dt} \langle \hat{P}_{\bf k} \rangle_R = -i
\tilde \omega_{\bf k} \langle \hat{P}_{\bf
k} \rangle_R + g_{\bf k} \langle \hat{P}^{\dag}_{\bf k_i} \rangle_R \mathcal{P}^2_{{\bf k}_p} \nonumber \\
\frac{d}{dt} \langle \hat{P}^\dag_{\bf k_i} \rangle_R = i
\tilde\omega_{\bf k_i} \langle \hat{P}^\dag_{\bf k_i} \rangle_R +
g^*_{\bf k_i} \langle \hat{P}_{\bf k} \rangle_R \mathcal{P}^2_{{\bf
k}_p}\, ,\end{eqnarray} where we changed slightly the notation to
underline that pump polariton amplitudes $\mathcal{P}_{{\bf k}_p}$
are regarded as classical variables (${\mathbb{C}}$-numbers), while
the generated signal and idler polaritons are regarded as true
quantum variables.

The nonlinear interaction terms in Eq.\, (\ref{sist lineare val
asp}) reads
\begin{equation}
g_{\bf k} = \frac{-i}{N_{eff}} \bigg[ \frac{V}{n_{sat}} C^*_{{\bf
k}_p} +  V_{\text{xx}} X^*_{{\bf k}_p} \bigg] X_{{\bf k}_p}
X^*_{{\bf k}_p} X_{{\bf k}_i}\, .
\end{equation}
It accounts for a pump-induced blue-shift of the polariton
resonances and a pump-induced parametric emission. In Eqs.\,
(\ref{sist lineare val asp}) only nonlinear terms arising from
saturation and from the mean-field Coulomb interaction have been
included. Correlation effects beyond mean-field introduce
non-instantaneous nonlinear terms. They mainly determine an
effective reduction of the mean-field interaction and an excitation
induced dephasing. It has been shown \cite{Savasta PRL2003} that
both effects depends on the sum of the energies of the scattered
polariton pairs. While the effective reduction can be taken into
account simply modifying $V_\text{xx}$, the proper inclusion of the
excitation induced dephasing requires the explicit inclusion into
the dynamics of four-particle states with their phonon-induced
scattering and relaxation. In the following we will neglect this
effect that is quite low at zero and even less at negative detuning
on the lower polariton branch \cite{Savasta PRB2001}. The
renormalized complex polariton dispersion $\tilde \omega_{\bf k}$
includes the effects of relaxation and pump-induced renormalization,
$ \tilde \omega_{\bf k} = \omega_{{\bf k}} -
i\Gamma^{(\text{tot})}_{\bf k}/{2} + h_{\bf k} \left|
\mathcal{P}_{{\bf k}_p} \right|^2\, , $ and
\begin{eqnarray}
h_{\bf k} = \frac{1}{N_{eff}} \big( \frac{V}{n_{sat}} C^*_{{\bf
k}_p} X_{{\bf k}_p} \left|X_{\bf k}\right|^2 + \frac{V}{n_{sat}}
C^*_{{\bf k}}X_{\bf k} \left|X_{{\bf k}_p}\right|^2\nonumber \\
+ 2 V_{\text{xx}}\left|X_{{\bf k}_p}\right|^2 \left|X_{{\bf
k}}\right|^2 \big)\, .
\end{eqnarray}
The damping term $\Gamma^{(\text{tot})}_{\bf k}$ here can be
regarded as a result of a microscopic calculation including a
thermal bath (see next Sect.).

Following Lax's prescription we can promote Eqs.\, (\ref{sist
lineare val asp}) to global bare-operator equations
\begin{eqnarray}\label{sist
lineare} \frac{d}{dt} \hat{P}_{\bf k} = -i \tilde \omega_{\bf
k}\hat{P}_{\bf
k} + g_{\bf k} \hat{P}^{\dag}_{\bf k_i} \mathcal{P}^2_{{\bf k}_p} + \hat{\mathcal{F}}_{\hat{P}_{\bf k}} \nonumber \\
\frac{d}{dt} \hat{P}^\dag_{\bf k_i} = i \tilde\omega_{\bf k_i}
\hat{P}^\dag_{\bf k_i} + g^*_{\bf k_i} \hat{P}_{\bf k}
\mathcal{P}^2_{{\bf k}_p} + \hat{\mathcal{F}}_{\hat{P}^\dag_{\bf
k_i}}\, .
\end{eqnarray}
However, in this form it is not a ready-to-use ingredient, indeed
its implementation in calculating spectra and/or higher order
correlators would be problematic because the noise commutation
relations ask for the solution of the same (at best of an analogous)
kinetic problem to be already at hand. This point can be very well
explained as soon as one is interested in calculating $\langle \hat
P^\dag_{{\bf k}} \hat P_{{\bf k}} \rangle $, i.e. the polariton
occupation, where the mere calculation is self-explanatory. We shall
need

\begin{equation}\label{Diff coeff} 2\langle \hat{D}_{P^\dag_{\bf k}
P_{\bf k}} \rangle_R = \frac{d}{d t} \langle \hat{P}^\dag_{\bf k}
\hat{P}_{\bf k} \rangle_R - \langle {\Big \{} \frac{d}{d t}
\hat{P}^\dag_{\bf k} {\Big \}}\hat{P}_{\bf k} \rangle_R - \langle
\hat{P}^\dag_{\bf k} {\Big \{} \frac{d}{d t} \hat{P}_{\bf k} {\Big
\}} \rangle_R\, ,\end{equation} and the diffusion coefficient for
the two operators in reverse order. Thanks to the structure above we
can easily see that all the coherent contributions cancel out and
only the incoherent ones are left. Anyway the important fact for the
present purpose is that they are proportional to the polaritonic
occupation, these coefficients will be explicitly calculated in
Section \ref{V}.

The general solution of Eqs.\, (\ref{sist lineare}) in the pump
reference frame reads:
\begin{eqnarray}\label{sistema}
&& \mathbf{P}(t) = e^{\int_0^t{\mathbf{\cal M} (t')}dt'}
\mathbf{P}(0) + \int_0^t e^{\int_{t'}^t\mathbf{\cal M}(t'')dt''}
 \mathbf{\cal K}(t')\, dt' \nonumber \\
&& \mathbf{P}(t) = \left( \begin{array}{c} \hat{P}_{\bf k}(t) \\
\hat{\overline{P}}^\dag_{\bf 2k_p - k}(t)
\end{array} \right),
 \mathbf{\cal K}= \left( \begin{array}{c}
{ \hat{\mathcal{F}}}_{\hat{P}_{s \bf k}} \\
{ \mathcal{\hat{\overline F}}}_{\hat{P}^\dag_{i \bf 2k_p - k}}
\end{array} \right) \\
&& \mathbf{\cal M}=\left( \begin{array}{cc} \overline \omega_{\bf
k}&  \Delta({\bf k},\tau)\\
\Delta^*({\bf k},\tau) &\overline \omega^*_{2{\bf k}_p -{\bf k}}
\end{array} \right) \nonumber\, ,\end{eqnarray}
where
\begin{eqnarray} && \overline{\omega}_{\bf k} = - i \tilde
\omega_{\bf k}\, , \nonumber \\
&& \overline{\omega}_{2{\bf k}_p -{\bf k}} = -
i(\tilde \omega_{2{\bf k}_p -{\bf k}} - 2 \tilde \omega_{{\bf k}_p})\, ,\nonumber \\
&& \text{the pump is}\ \ \mathcal{P}_{\bf k_p} = \mathcal{P}^o_{\bf
k_p} e^{-i\omega^{(h)}_{\bf k_p} t}\, ,\nonumber \\
&&\hat{\overline{P}}^\dag_{2 \bf{k_p - k}} = \hat{P}^\dag_{ 2
\bf{k_p - k}} e^{-i 2 \omega^{(h)}_{\bf{k_p}}t}\, ,\nonumber \\
&&  \hat{\mathcal{\overline{F}}}_{\hat{P}^\dag_{2 \bf{k_p - k}}} =
\hat{\mathcal{F}}_{\hat{P}^\dag_{ 2 \bf{k_p - k}}}
e^{-i 2\omega^{(h)}_{\bf{k_p}}t}\, ,\nonumber \\
&& \omega^{(h)}_{\bf k} = \omega_{{\bf k}} + h_{\bf k} \left|\mathcal{P}_{\bf k_p} \right|^2 \, ,\nonumber \\
&& \Delta({\bf k},\tau) = g_{\bf k} \mathcal{P}^{o\ 2}_{\bf k_p}\,
.\end{eqnarray}

Equation (\ref{sistema}) can be written in a more explicit form by
exploiting the following identity:
\begin{eqnarray}\label{metodo sys}
&& e^{\int^{t_2}_{t_1}A(t)dt} = \alpha_1(t1,t2)
\int^{t_2}_{t_1}A(t)dt + \alpha_0(t1,t2)\mathbb{I} \nonumber \\
&& \alpha_0(t1,t2) = \frac{\Lambda_+(t1,t2)e^{\Lambda_-(t1,t2)} -
\Lambda_-(t1,t2)e^{\Lambda_+(t1,t2)}}{\Lambda_+(t1,t2) -
\Lambda_-(t1,t2)} \nonumber \\
&& \alpha_1(t1,t2) = \frac{e^{\Lambda_+(t1,t2)} -
e^{\Lambda_-(t1,t2)}}{\Lambda_+(t1,t2) - \Lambda_-(t1,t2)} \nonumber \\
&& \Lambda_{\pm}(t1,t2) = \int^{t_2}_{t_1}\lambda_{\pm}(\tau)d\tau
\nonumber \\
&& \int^{t_2}_{t_1}\text{diag}[A(t)]dt = \left( \begin{array}{cc}
\Lambda_-(t1,t2) &0 \\
0 &\Lambda_+(t1,t2)\end{array} \right) \nonumber \\
&& \lambda_{\pm} = w^+ \pm \sqrt{(w^-)^2 + |\Delta|^2} \nonumber \\
&& w^+ = \frac{(\overline{\omega}_{\bf k} +
\overline{\omega}^*_{2{\bf k}_p -{\bf k}})}{2}, \ \ w^- =
\frac{(\overline{\omega}_{\bf k} - \overline{\omega}^*_{2{\bf k}_p
-{\bf k}})}{2}\, .\end{eqnarray} Eq.\, (\ref{sistema}) with Eq.\,
(\ref{metodo sys}) provides an easy and general starting point for
the calculation of multi-time correlation functions which are
key-quantities in quantum optics. Taking the expectation values of
the appropriate products it yields
\begin{eqnarray}\label{PcrP Langevin formal}
&& \langle \hat P^\dag_{{\bf k}} \hat P_{{\bf k}} \rangle =
|c_1(0,t)|^2 \langle \hat P^\dag_{{\bf k}} \hat P_{{\bf k}}
\rangle(0) + |c_2(0,t)|^2 \langle \hat P_{{\bf k}} \hat P^\dag_{{\bf k}} \rangle(0) + \nonumber \\
&& + \int^{t}_0 d\tau |c_1(\tau,t)|^2 2\langle \hat{D}_{P^\dag_{\bf
k} P_{\bf k}} \rangle (\tau) + \int^{t}_0 d\tau |c_2(\tau,t)|^2
2\langle \hat{D}_{P_{\bf k} P^\dag_{\bf k}} \rangle \,
,\end{eqnarray} here
\begin{eqnarray} && c_1(t_1,t_2) =\alpha_1(t1,t2) \int^{t_2}_{t_1}
d\tau (-\frac{\Gamma_{\bf k}^{(tot)}}{2} - i\omega^{(h)}_{\bf k}) +
\alpha_0(t1,t2)\nonumber \\
&& c_2(t_1,t_2) =\alpha_1(t1,t2) \int^{t_2}_{t_1} d\tau \Delta({\bf
k},\tau)\, .\end{eqnarray}
The two diffusion coefficients are proportional to the polariton
occupation, i.e. we need as known input sources the very quantities
we are about to calculate and a self-consistent solution seems
unavoidable. Concluding, even if exact, Lax's theorem is of no
immediate use for it simply rearranges the various ingredients to
the microscopic dynamics in a different way. It seems worth noticing
however that what up to now appears as a very formal and academic
line of reasoning will be the clue for all the subsequent physical
arguments ending up into an innovative approach to quantum optics in
the strong coupling regime. Indeed, as we shall see in Sect.\,
\ref{V}, under certain assumptions we will be able to overcome the
above mentioned difficulty elaborating a (computationally
advantageous) decoupling of incoherent dynamics from  parametric
processes.

Anyway, it is the structure of Eq.\, (\ref{Diff coeff}) for the
diffusion coefficients which allows, physically speaking, to account
for each contribution in its best proper way recognizing easily the
dominant contribution. Indeed, reconsidering Eq.\, (\ref{Diff
coeff}) in the light of the proper kinetic equation for the
polariton population dynamics --- the subject of the following
section ---, it is very clear that thanks to its structure all the
coherent contributions cancel out automatically giving us an easy
way to separate coherent and incoherent parts, but at the same time
to treat them on an equal footing when calculating the final result.

\section{Microscopic Markov calculation of Polariton
Photoluminescence}\label{IV}

Excitonic polaritons  propagate in a complex interacting environment
and contain real electronic excitations subject to scattering events
and noise, mainly originating from the interaction with lattice
vibrations, affecting quantum coherence and entanglement. For a
realistic description of the physics in action, we need to build up
a microscopic model taking into account on an equal footing
nonlinear interactions, light quantization, cavity losses and
polariton-phonon interaction. To be more specific as a dominant
process for excitonic decoherence in resonant emission from QWs we
shall consider acoustic-phonon scattering via deformation potential
interaction, whereas we shall model the losses through the cavity
mirrors within the quasi-mode approach (see Appendix\,
\ref{exc2pol}). It is worth pointing out that the approach we are
proposing may be easily enriched by several other scattering
mechanisms suitable for a refinement of the numerical results.

In the view of the change of bases previously-mentioned, so
imperative for a proper Markov calculation, we decide to treat the
coupled system, described by the three Hamiltonian terms
$\hat{H}_e$, $\hat{H}_c$ and $\hat{H}_I$, as our system of interest
weakly interacting with the environment. In practice, this means to
start from the linear part of the Heisenberg equations of motion in
Eq.\, (\ref{2}), which can be considered in the spirit of Sect.\,
\ref{III} as system-operator equations, without the input term. Once
obtained the polariton modes via a unitary diagonalizing
transformation Eq.\, (\ref{diagonalization}), we apply, to the
coupling of this system with the environment, the usual many-body
perturbative description. We end up with the customary
Bogoliubov-Born-Green-Kirkwood-Yvon (BBGKY) hierarchy which to the
first order gives us the coherent input field, whereas to the second
order the phonon and radiative scattering terms. As widely used in
the literature, we shall limit ourselves up to this point thus
performing a second-order Born-Markov description of the environment
induced effects to the system dynamics. To exemplify our approach we
shall calculate the relaxation rate of  $\langle \hat{O} \rangle =
\langle\hat P^\dag_{\alpha {\overline{\bf k}}} \hat P_{\alpha
\overline{\bf k}}\rangle$ in the sole case of acoustic phonon
interaction, any other scattering mechanism will be treated in the
same way. The rate equation governing the incoherent dynamics to the
lowest order of the polariton occupation is a relevant quantity we
exploit in the next Sect. when we propose our DCTS-Langevin recipes
for the calculation of multi-time many-body correlation functions.
Being the full trace of the polariton density over the reservoir and
the system degrees of freedom a relevant physical observable that we
need to solve numerically, we prefer to give explicitly full account
of the manipulations we have followed. It is worth underline that
the very same formal treatment, i.e. Markov approximation, can be
performed easily on the system operator arisen form the partial
trace over the reservoir density matrix, $\langle \ \ \rangle_R$
\cite{Lax}. In this latter guise damping and dephasing enter the
{\emph mean system operator} equation (\ref{lax mean eq}), starting
point for the Lax's theory of quantum noise \cite{Lax}. A completely
analogous procedure can be followed for the calculation of the
dephasing rate of $\langle \hat P_{\alpha \overline{\bf k}}
\rangle$.

In the following the DCTS description of the interaction with the
environment is limited to the lowest order. This means that effects
like final state stimulation of scattering events are neglected. At
the lowest order, the acoustic phonon interaction Hamiltonian can be
expressed only in terms of excitonic operators as \cite{Takagahara}
\begin{eqnarray} H^{DF}_{exc-ph}\hspace{-0.1cm} =&& \hspace{-0.3cm} \sum_{\bf q}
\hspace{-0.1cm} \left(\sum_{\bf k} t^{\bf q}_{\bf k}\ket{1 S \, {\bf
k+q^\|}} \bra{1 S\, {\bf k}} \right) \left( b_{\bf q} +
b^\dag_{-{\bf
q}} \right) = \nonumber \\
&& \hspace{-0.3cm}= \sum_{\bf q} Q_{\bf q} F_{\bf q} + Q_{\bf q}
F^\dag_{- \bf q}\, ,
\end{eqnarray} $t^{\bf q}_{\bf k}$ is described in the Eq.\, (\ref{t coeff})
of the Appendix. Simbolically $\hat{H}^S$ stands for all the system
Hamiltonians, i.e. free dynamics and parametric scattering. The
standard microscopic perturbative calculation \cite{Lax} gives:
\begin{eqnarray}\label{Markov AP}
&&\hspace{-1.0cm} \frac{d}{dt}\langle \hat{O}\rangle= \left<\right.
\frac{1}{i \hbar} \left[ \hat{O},\hat{H}^S \right] \left.\right>
\hspace{0.0cm} - \frac{1}{\hbar^2} \sum_{\bf q, \pm} \int^{\infty}_0 du \left( n^R_{\mp \bf q} + \theta(\pm) \right)\nonumber\\
&&\times {\Big (} e^{\pm \frac{\epsilon^R_{\bf q} u}{i \hbar}}
\left<\right. \left.\right[ \hat{O}, \hat{Q}_{\bf q} \left.\right]
\hat{Q}_{- \bf q}(-u) \left.\right> - e^{\mp \frac{\epsilon^R_{- \bf
q} u}{i \hbar}} \left<\right. \hat{Q}_{- \bf q}(-u) \left.\right[
\hat{O}, \hat{Q}_{\bf q} \left.\right] \left.\right> {\Big )}\, ,
\end{eqnarray} where the meaning of the new symbols are
self-evident.

Within the strong coupling region, the dressing carried by the
nonperturbative coupling between excitons and cavity photons highly
affects the scattering and for a microscopic calculation we are
urged to leave the couple picture of Eqs.\, (\ref{2}) and move our
steps into the polaritonic operator bases. Our aim is to produce a
microscopic description of damping and fluctuation and to apply it
in experiments with low and-or moderate excitation intensities, thus
we expect the strong-coupling regime to become crucial in the
scattering rates mainly through the polaritonic spectrum. In the
spirit of the DCTS we shall consider them as transitions over the
polaritonic bases obtained form excitons and cavity modes states and
the linear diagonalizing transformation Eq.\,
(\ref{diagonalization}), form exciton and photon operators to
polariton operators, can then be rewritten as
\begin{equation} \ket{1 S\,{\bf k}'} \bra{1 S\,{\bf k}} = \sum_{i,j}
X_{i {\bf k}'} X^*_{j {\bf k}} \ket{i {\bf k}'} \bra{j {\bf k}}\, ,
\end{equation} where it is understood we have transition operators on the left-hand side
representing excitons, whereas on the right-hand side polaritons.
Within the Born-Markov description we are left with

\begin{equation}
\frac{d}{d t} \langle \hat P^\dag_{\alpha \overline{{\bf k}}} \hat
P_{\alpha \overline{{\bf k}}}
\rangle_{\left|_{H^{DF}_{exc-ph}}\right.} = -
\Gamma^{ph}_{\alpha,\overline{{\bf k}}} \langle \hat P^\dag_{\alpha
\overline{{\bf k}}} \hat P_{\alpha \overline{{\bf k}}} \rangle +
\sum_{l {\bf k'}} W_{(\alpha \overline{{\bf k}}),(l {\bf k'})}
\langle \hat P^\dag_{l {\bf k'}} \hat P_{l {\bf k'}} \rangle\, ,
\end{equation}
with

\begin{eqnarray}
&& W^{\pm}_{(s {\bf k}),(r {\bf k'})} = \frac{2 \pi}{\hbar}
\sum_{q_z} \left|\right. t^{{\bf k}',q_z}_{\bf k} \left.\right|^2
\left|\right. X_{s {\bf k}} \left.\right|^2 \left|\right. X_{r {\bf
k}'} \left.\right|^2 \nonumber \\
&& \hspace{4.0cm} \delta(\epsilon_{s {\bf k}} - \epsilon_{r{\bf k}'}
\pm \hbar \omega^R_{({\bf k}' - {\bf k},q_z)}) \left(\right.
n_{({\bf k}' - {\bf k},q_z)} + \frac{1}{2} \pm \frac{1}{2}
\left.\right)
\nonumber \\
&& W_{(s {\bf k}),(r {\bf k'})} = \sum_{\pm} W^{\pm}_{(s {\bf k}),(r
{\bf k'})} \nonumber \\
&& \Gamma^{(ph)}_{s,{\bf k}} = \sum_{r {\bf k}'} W_{(r {\bf k'}),(s
{\bf k})}\, .
\end{eqnarray}
These happen to be the same ingredients used in Ref.\,
\cite{Piermarocchi bottleneck} studying bottleneck effect in
relaxation and photoluminescence of microcavity polaritons within a
bosonic Boltzmann approach.

Within the quasi-mode approach, the emitted light is proportional to
the intracavity photon number ($t_c$ is the transmission
coefficient):
\begin{equation}\label{PL(t) def}
I_{\bf k}^{PL}(t) = t_c^2 \langle \hat{a}^\dag_{\bf k} \hat{a}_{\bf
k} \rangle(t) = t_c^2 \sum_{i} |C_{i {\bf k}}|^2 \langle
\hat{P}^\dag_{i {\bf k}} \hat{P}_{i {\bf k}} \rangle(t)\, .
\end{equation}
By applying the whole machinery, when including only the lowest
order terms in the input light field, the following equation for the
polariton-occupations dynamics is obtained:
\begin{eqnarray}\label{Pcroce P}
\frac{d}{d t} \hspace{-0.4cm} && \langle \hat P^\dag_{i {\bf k}_i}
\hat P_{i {\bf k}_i} \rangle = - (\Gamma^{ph}_{i,{\bf k}_i} +
\gamma^{(c)}_{i,{\bf k}_i})
\langle \hat P^\dag_{i {\bf k}_i} \hat P_{i {\bf k}_i} \rangle  \\
&&  + g_{i,{\bf k}_i} + \Gamma^{(c)}_{i,{\bf k}_i} + \sum_{l {\bf
k'}} W_{(i {\bf k}_i),(l {\bf k'})} \langle \hat P^\dag_{l {\bf k'}}
\hat P_{l {\bf k'}} \rangle \nonumber\, ,
\end{eqnarray}

with the generation rate given by \begin{equation}\label{gen rate}
g_{i,\bf k} = {\Big [} t_c C_{{\bf k}_p} E^{in(+)}_{{\bf k}}
\delta_{{\bf k},{\bf k}_p} \langle \hat{P}^\dag_{i,{\bf k}}
\rangle_{coh} + t_c C^*_{{\bf k}_p} E^{in(-)}_{{\bf k}}\delta_{{\bf
k},{\bf k}_p} \langle \hat{P}_{i,{\bf k}} \rangle_{coh} {\Big ]}\, .
\end{equation}
The phonon-emission $(+)$ and phonon-absorbtion $(-)$ scattering
rates read
\begin{eqnarray}
W^{\pm}_{(j {\bf k}'),(i {\bf k})}&& \hspace{-0.5cm} = \frac{1}{\rho
u S} \frac{|k'-k|^2 +(q^0_z)2}{|\hbar u q^0_z|} |\Xi|^2 \\
&& \times |X_{j \bf k'}|^2 |X_{i \bf k}|^2
[n^{(ph)}(E^{ph}_{(k'-k,q^0_z)}) + \frac{1}{2} \pm \frac{1}{2}]
\nonumber\, ,
\end{eqnarray} the 3D phonon wave vector is $({\bf
q},q^0_z)$, whereas $q^0_z$ is calculated so that the energy
conservation delta function $\delta(\hbar \omega_{j {k'}} - \hbar
\omega_{i {k}} \pm E^{ph}_{\bf q})$ is satisfied,
\begin{equation}
\Xi = \left( D_c I^\bot_e(q_z) I^\|_e({\bf k'-k}) - D_v
I^\bot_h(q_z) I^\|_h({\bf k'-k}) \right)\, , \end{equation} with the
overlap integrals
\begin{eqnarray}\label{over} && I^{\|}_{e(h)} ({\bf
q}) = \left[ 1 + \left( \frac{m_{ h(e)}}{2(m_e+m_h)}|{\bf q}|
a_{\text{x}}\right)2 \right]^{-3/2},\nonumber \\
&& I^{\bot}_{e(h)}(q_z) = \int_L dz |\chi_{e(h)}(z)|^2e^{iq_z z}\, .
\end{eqnarray} We shall treat the cavity field in the quasi-mode
approximation, that is to say we shall quantize the field as the
mirror were perfect and subsequently we shall couple the cavity with
a statistical reservoir of a continuum of external modes. This way
on an equal footing we shall provide the input coherent driving
mechanism (at first order in the interaction) and the radiative
damping channel (within a second order Born-Markov description).

The escape rate through the two mirrors ($l\equiv$left,
$r\equiv$right) is
\begin{equation}\label{gamma rad}
\gamma^{(c)}_{i,{\bf k^\|}} = \frac{2 \pi}{\hbar} |C_{i,{\bf
k^\|}}|^2 \sum_{s=l,r} \!\!\! \int d\omega
\delta\left(\Omega^{qm}_{\bf k^\|}[\omega] - \omega_{k^\|}\right)
\hbar |g_{s \bf k^\parallel}(\omega)|^2\, ,
\end{equation} and the corresponding noise term reads
\begin{equation}\label{Gamma rad}
\Gamma^{(c)}_{i,{\bf k^\|}} = \frac{2 \pi}{\hbar} |C_{i,{\bf
k^\|}}|^2 \sum_{s=l,r} \!\!\! \int d\omega n^{qm}_{\bf k^\|}(\omega)
\delta\left(\Omega^{qm}_{\bf k^\|}[\omega] - \omega_{k^\|}\right)
\hbar |g_{s \bf k^\parallel}(\omega)|^2\, .
\end{equation}

\section{DCTS-Nonequilibriun Quantum Langevin approach to parametric
emission}\label{V}

As already discussed, Eq.\, (\ref{sistema}) with Eq.\, (\ref{metodo
sys}) provides an easy and general starting point for the
calculation of multi-time correlation functions which are
key-quantities in quantum optics. Thus it would be easy-tempting to
wonder if, through some appropriate, thought and motivated physical
considerations, we were given the noise sources as known inputs.
Thanks to the structure of Eq.\, (\ref{Diff coeff}) for the
diffusion coefficients we are allowed, physically speaking, to
recognize properly the dominant contribution. In order to be more
\textit{specific} let us fix our attention on the explicit form of
$2\langle \hat{D}_{P^\dag_{\bf k} P_{\bf k}} \rangle$:
\begin{equation}\label{D PcrP} 2\langle \hat{D}_{P^\dag_{\bf k}
P_{\bf k}} \rangle (t) = \sum_{\bf k'} W_{{\bf k},{\bf k}'} \langle
\hat{P}^\dag_{\bf k'} \hat{P}_{\bf k'} \rangle (t) + \Gamma^{c}_{\bf
k}\, . \end{equation}
Inspecting Eq.\, (\ref{Pcroce P}), it results that in the low and
intermediate excitation regime the main incoherent contribution to
the dynamics is the PL the pump produces by itself, the effects on
the PL of subsequent pump-induced repopulation arising from the
nonlinear parametric part is negligible, that is to say the
occupancies of the couple signal-idler are at least one order of
magnitude smaller than the pump occupancy. This means that in Eq.\,
(\ref{D PcrP}) we can consider at the right hand side the solution
in time of Eq.\, (\ref{Pcroce P}), i.e. only incoherent lowest order
contributions. The other important diffusion coefficient reads:
\begin{eqnarray}\label{D PPcr} 2\langle \hat{D}_{P_{\bf k}
P^\dag_{\bf k}} \rangle (t) \hspace{-0.3cm} &&= \sum_{\bf k'}
W_{{\bf k},{\bf k}'} \langle \hat{P}_{\bf k'} \hat{P}^\dag_{\bf k'}%
\rangle (t) +
\Gamma^{c}_{\bf k} + \gamma^{c}_{\bf k} =  \nonumber \\
&& \hspace{-1.0cm} = \sum_{\bf k'} W_{{\bf k},{\bf k}'}
\left(\right. \langle \hat{P}^\dag_{\bf k'} \hat{P}_{\bf k'} \rangle
(t) + 1 \left.\right) + \Gamma^{c}_{\bf k} + \gamma^{c}_{\bf k}\,
.\end{eqnarray}
We decide to use a sort of bosonic-like commutation relation in the
equation above but only in the present situation, restricted only to
this precise case and to the noisy background $\langle \hat{P}_{\bf k} \hat{P}^\dag_{\bf k}%
\rangle(0)$, responsible for spontaneous parametric emission. The
reason is many-fold. The two terms, Eq.\, (\ref{D PPcr}) and the
above noise background, will enter in Eq.\, (\ref{PcrP Langevin
formal}) multiplied by $|c_2|^2$ which contains the pump already
twice. As a consequence their contribution must be to linear order.
Besides, Eq.\, (\ref{Pcroce P}) can be considered as the very low
density limit of the rate equation obtained from the picture of
polaritons as bosons as obtained in Ref.\, \cite{Piermarocchi
bottleneck}. It witnesses that when the focus is devoted to the sole
incoherent lowest order dynamics, bosonic commutation rules for
polaritons may be employed, though carefully. Moreover, direct
computation for normal incidence gives
\begin{equation}\label{[]}
[\hat{B}_{n }, \hat{B}^\dag_{n'} ] = \delta_{n',n} - \sum_{\bf q}
\Phi^{*}_{n {\bf q}} \Phi^{}_{n' {\bf q}} \sum_{N,\alpha,\beta}
\Biggl(\bra{N \alpha}\hat{c}^\dag_{{\bf q}} c_{{\bf q}}\ket{N \beta}
+ \bra{N \alpha}\hat{d}^\dag_{-{\bf q}} d_{-{\bf q}}\ket{N \beta}
\Biggr)  \ket{N \alpha} \bra{N \beta}\, ,
\end{equation}
where as usual $\ket{N \beta}$ and $\ket{N \alpha}$ are N-pair {\em
eh} pairs. Thus, within a DCTS analysis, the dominant term to the
lowest order in the commutator is $\delta$-like, whereas the two
additional contributions, being proportional to the electron and
hole densities, are nonlinear higher order corrections, contributing
to the lowest order nonlinear dynamics but negligible for very low
density, i.e. linear order. In the following we shall indicate as
$\langle \hat{P}^\dag_{\bf k} \hat{P}_{\bf k} \rangle_{\text{PL}}$
this solution representing the (incoherent) polariton occupation of
the pump-induced PL. In the following subsection we show that this
choice guarantees consistency between the rate equation Eq.\,
(\ref{Pcroce P}) and the complete solution we are about to present
in Eq.\, (\ref{PcrP Langevin FINAL}) in the limit of pump intensity
tending to zero, i.e. when it is the incoherent PL which governs the
dynamics.

\subsubsection*{polariton occupation dynamics}

With the notation introduced so far, the Heseinberg-Langevin
equation governing the dynamics are those of Eq.\, (\ref{sist
lineare}) which we report here again:

\begin{eqnarray}\label{sist
lineare2} \frac{d}{dt} \hat{P}_{\bf k} = -i \tilde \omega_{\bf
k}\hat{P}_{\bf
k} + g_{\bf k} \hat{P}^{\dag}_{\bf k_i} \mathcal{P}^2_{{\bf k}_p} + \hat{\mathcal{F}}_{\hat{P}_{\bf k}} \nonumber \\
\frac{d}{dt} \hat{P}^\dag_{\bf k_i} = i \tilde\omega_{\bf k_i}
\hat{P}^\dag_{\bf k_i} + g^*_{\bf k_i} \hat{P}_{\bf k}
\mathcal{P}^2_{{\bf k}_p} + \hat{\mathcal{F}}_{\hat{P}^\dag_{\bf
k_i}}\, ,
\end{eqnarray}
where $ \tilde \omega_{\bf k} = \omega_{{\bf k}} -
i\Gamma^{(\text{tot})}_{\bf k}/{2} + h_{\bf k} \left|
\mathcal{P}_{{\bf k}_p} \right|^2\, , $ with
$\Gamma^{(\text{tot})}_{\bf k} = (\Gamma^{(ph)}_{\bf k} +
\gamma^{(c)}_{\bf k})$.

The general solution for the polariton occupation reads

\begin{eqnarray}\label{PcrP Langevin FINAL}
&& \langle \hat P^\dag_{{\bf k}} \hat P_{{\bf k}} \rangle =
|c_1(0,t)|^2 N_{\bf k}(0) + |c_2(0,t)|^2 (N_{2{\bf k}_p - {\bf
k}}(0) + 1) + \nonumber \\
&& + \int^{t}_0 d\tau |c_1(\tau,t)|^2 \sum_{\bf k'} W_{{\bf
k},{\bf k}'} \langle \hat{P}^\dag_{\bf k'} \hat{P}_{\bf k'} \rangle_{\text{PL}} (\tau) + \\
&& + \int^{t}_0 d\tau |c_2(\tau,t)|^2 \left( \sum_{\bf k'} W_{{2{\bf
k}_p - {\bf k}},{\bf k}'} \left( \langle \hat{P}^\dag_{\bf k'}
\hat{P}_{\bf k'} \rangle_{\text{PL}} (\tau) + 1 \right) +
\gamma^{(c)}_{{2{\bf k}_p - {\bf k}}} \right) \nonumber\,
.\end{eqnarray}
In all the situations under investigation, the thermal population of
photons at optical frequencies are negligible, hence
$\Gamma^{(c)}_{\bf k} \simeq 0$. Moreover, in the limit of pump
intensity tending to zero it is the PL which governs the dynamics.
Indeed Eq.\, (\ref{Pcroce P}) in this situation reads
\begin{equation} \frac{d}{d t} \langle \hat P^\dag_{\bf k} \hat
P_{\bf k} \rangle = - \Gamma^{(\text{tot})}_{\bf k} + \langle \hat
P^\dag_{\bf k} \hat P_{\bf k} \rangle + \sum_{\bf k'} W_{{\bf
k},{\bf k'}} \langle \hat P^\dag_{\bf k'} \hat P_{\bf k'} \rangle
\nonumber\, ,
\end{equation}when, at least formally, we consider the right-hand side as known we
can integrate, obtaining \begin{equation} \langle \hat P^\dag_{\bf
k} \hat P_{\bf k} \rangle = \int^t_0 dt' e^{-
\Gamma^{(\text{tot})}_{\bf k} (t-t')} \sum_{\bf k'} W_{{\bf k},{\bf
k'}} \langle \hat P^\dag_{\bf k'} \hat P_{\bf k'} \rangle \nonumber
\end{equation}which is the limit of excitation intensity to zero of
Eq.\, (\ref{PcrP Langevin FINAL}). The form of $2\langle
\hat{D}_{P_{\bf k} P^\dag_{\bf k}} \rangle$ guarantees this fact for
the reverse order calculation.

In Eq.\, (\ref{PcrP Langevin FINAL}) it is evident the great
flexibility of the Langevin method, even in single-time
correlations. It represents a clear way to ``decouple" the
incoherent and the coherent dynamics in an easy and controllable
fashion. In the important case of steady-state, where the standard
Langevin theory could at least in principle be applied, we have
nonequilibrium Langevin sources which become:
\begin{eqnarray}\label{ss coeff}
&& 0 =  - \Gamma^{(\text{tot})}_{\bf k} \langle \hat P^\dag_{\bf k}
\hat P_{\bf k} \rangle + \sum_{\bf k'} W_{{\bf k},{\bf k'}} \langle
\hat P^\dag_{\bf k'} \hat P_{\bf k'} \rangle + \Gamma^{c}_{\bf k} \nonumber \\
&& 2\langle \hat{D}_{P^\dag_{\bf k} P_{\bf k}} \rangle = \sum_{\bf
k'} W_{{\bf k},{\bf k}'} \langle \hat{P}^\dag_{\bf k'} \hat{P}_{\bf
k'} \rangle (t) + \Gamma^{c}_{\bf k} \nonumber \\
&& 0 =  - \Gamma^{(\text{tot})}_{\bf k} \langle \hat P_{\bf k} \hat
P^\dag_{\bf k} \rangle + \sum_{\bf k'} W_{{\bf k},{\bf k'}} \langle
\hat P_{\bf k} \hat P^\dag_{\bf k} \rangle+ \Gamma^{c}_{\bf k} +
\gamma^{c}_{\bf k} \nonumber \\
&& 2\langle \hat{D}_{P_{\bf k} P^\dag_{\bf k}} \rangle= \sum_{\bf
k'} W_{{\bf k},{\bf k}'} \langle \hat P_{\bf k} \hat P^\dag_{\bf k}
\rangle + \Gamma^{c}_{\bf k} + \gamma^{c}_{\bf k} \nonumber\, ,
\end{eqnarray} giving \begin{eqnarray}\label{ss coeffFIN} && 2\langle \hat{D}_{P^\dag_{\bf k}
P_{\bf k}} \rangle (t) = \Gamma^{(\text{tot})}_{\bf k} \langle
\hat{P}^\dag_{\bf k} \hat{P}_{\bf k}
\rangle (t)\nonumber \\
&& 2\langle \hat{D}_{P_{\bf k} P^\dag_{\bf k}} \rangle (t) =
\Gamma^{(\text{tot})}_{\bf k} (\langle \hat{P}^\dag_{\bf k}
\hat{P}_{\bf k} \rangle  (t) + 1) \nonumber\, ,
\end{eqnarray} i.e. the standard statistical viewpoint is recovered in
steady-state.

Concluding this section, standard Langevin theory gives some
problems in dealing with interaction forms more complicated than the
standard linear two-body coupling and some additional approximations
are needed. Lax technique, on the contrary, provides us with the
correct Langevin noise sources in the generic nonequilibrium case no
matter of the operatorial form of the reservoir (weak) interaction
Hamiltonian to implement. They properly recover the well-known
steady-state result even if they depend, in the generic case, on the
scattering rates rather than on dampings, on the contrary to
standard Langevin description.

\subsubsection*{time-integrated spectrum}
The spectrum of a general light field has always been of great
interest in understanding the physical properties of light. Any
spectral measurement is made by inserting a frequency-sensitive
device, usually a tunable linear filter, in front of the detector.
What is generally called ``spectrum" of light is just an
\textit{appropriately normalized record of the detected signal as a
function of the frequency setting of the filter}\cite{Eberly}. Here
we are interested to the power spectrum of a quantum-field
originating from pulsed excitation and thus not at steady state. The
time-integrated spectrum of light for a quantum-field can be
expressed as \cite{Cresser,Eberly}.
\begin{equation}\label{def1}
\mathcal{I}_{\bf k} (\omega, T) = \kappa \frac{2 \Gamma}{T-t_0}
\int^T_{t_0} dt_1 \int^T_{t_0} dt_2 \langle \hat{E}^{(-)}_{\bf
k}(t_1) \hat{E}^{(+)}_{\bf k}(t_2) \rangle e^{-(\Gamma -i \omega) (T
- t_1)} e^{-(\Gamma +i \omega) (T - t_2)}\, ,
\end{equation}
where $\Gamma$ is the bandwidth of the spectrometer (e.g. of the
Fabry-Perot interferometer) and $\hat{E}^{(-)}_{\bf k}$ ($
\hat{E}^{(+)}_{\bf k}$) are the field operators corresponding to the
light impinging on the detector, $\kappa$ is nothing but a
proportional factor depending on the detector parameters and
efficiency. Within the quasi-mode approach \cite{Collet Gardiner PRA
1985} the spectrum of transmitted light  is proportional to the
spectrum of the intracavity field. In our situation, in the very
narrow bandwidth limit and considering a beam with given in-plane
wave vector, \cite{Renaud} it reads
\begin{equation}\label{def2}
\mathcal{I}_{\bf k} (\omega, T) = \frac{\kappa \ {t_c}^2}{T-t_0}
\int^T_{t_0} dt_1 \int^T_{t_0} dt_2 \langle \hat{a}^\dag_{\bf
k}(t_1) \hat{a}_{\bf k}(t_2) \rangle e^{-i \omega (t_1 - t_2)}\, .
\end{equation}
By expressing the cavity-photon operator in terms of polariton
operators, one obtains
\begin{equation}\label{definition}
\mathcal{I}_{\bf k} (\omega, T) = \frac{\kappa \ {t_c}^2}{T-t_0}
\int^T_{t_0} dt_1 \int^T_{t_0} dt_2 \sum_i |C_{i {\bf k}}|^2 \langle
\hat{P}^\dag_{i \bf k}(t_1) \hat{P}_{i \bf k}(t_2) \rangle e^{-i
\omega (t_1 - t_2)}\, .
\end{equation}
In our experimental conditions the upper polariton contribution is
negligible, thus we need to calculate
\begin{equation}\label{da calcolare}
\mathcal{I}_{\bf k} (\omega, T) = \frac{\kappa \ {t_c}^2 |C_{\bf
k}|^2}{T-t_0} \int^T_{t_0} dt_1 \int^T_{t_0} dt_2 \langle
\hat{P}^\dag_{\bf k}(t_1) \hat{P}_{\bf k}(t_2) \rangle e^{-i \omega
(t_1 - t_2)}\, .
\end{equation}
By using Eq.\, (\ref{sistema}) and the properties of noise operators
(\ref{Fluc-Diss}), one obtains:
\begin{eqnarray}\label{t1 t2 uno}
&& \langle \hat P^\dag_{{\bf k}}(t_1) \hat P_{{\bf k}}(t_2) \rangle
= c_1(0,t_1)^* c_1(0,t_2) N_{\bf k}(0) + c_2(0,t_1)^* c_2(0,t_2)
(N_{2{\bf k}_p - {\bf
k}}(0) + 1) + \nonumber \\
&& + \delta_{t_{\alpha},\text{min}(t_1,t_2)} \int^{t_{\alpha}}_0
d\tau \ c_1(\tau,t_1)^* c_1(\tau,t_2) \sum_{\bf k'} W_{{\bf
k},{\bf k}'} \langle \hat{P}^\dag_{\bf k'} \hat{P}_{\bf k'} \rangle_{\text{PL}} (\tau) + \\
&& + \delta_{t_{\alpha},\text{min}(t_1,t_2)} \int^{t_{\alpha}}_0
d\tau \ c_2(\tau,t_1)^* c_2(\tau,t_2) \left( \sum_{\bf k'} W_{{2{\bf
k}_p - {\bf k}},{\bf k}'} \left( \langle \hat{P}^\dag_{\bf k'}
\hat{P}_{\bf k'} \rangle_{\text{PL}} (\tau) + 1 \right) +
\gamma^{(c)}_{{2{\bf k}_p - {\bf k}}} \right) \nonumber\,
.\end{eqnarray}

\section{Numerical Results}\label{VI}

In order to perform numerical calculations, we need to discretize in
${\bf k}$-space. Although, thanks to confinement, cavity photons
acquire a mass, it is about 4 order of magnitude smaller than the
typical exciton mass, thus the polariton splitting results in a very
steep energy dependence on the in-plane wave vector  near ${\bf
k}=0$ ($k= \omega \sin{\theta}/c$) (see Fig.\, 1). This very strong
variation of the polariton effective mass with momentum makes
difficult the numerical integration of the polariton PL
rate-equations (\ref{Pcroce P}). For example if PL originates from a
pump beam set at the {\em magic-angle} (see Fig.\, 1) or beyond, a
small temperature of $5$ K is sufficient to enable scattering
processes towards states at quite higher $k$-vectors, thus it is
necessary to include a computational window in k-space,
significantly beyond ${\bf k}_{pump}$. Usually, in finite volume
numerical calculations, the ${\bf k}$-space mesh is chosen  uniform,
but a dense grid suitable for the strong coupling region would
result in a grid of prohibitively large number of points for (e.g.
thermally activated) higher ${\bf k}$, on the other hand a mesh
well-suited for polaritons at higher k-values would consist of so
few points close to ${\bf k}=0$ to spoil the results gathered from
the numerical code completely of their physical significance.
Following Ref.\ \cite{Tassone Yam PRB 1999} we choose a uniformly
spaced grid in energy which results in an adaptive ${\bf k}$-grid
(in modulus), in addition, thanks to the rotating symmetry of the
dispersion curve, we choose a uniformly distributed mesh in the
angle $\theta$ so that ${\bf k}=(k,\theta)$. Unfortunately,  even if
this choice allows for a numerical integration of the polariton PL
rate equations (\ref{Pcroce P}), it provides an unbearable poor
description of the parametric processes (\ref{PcrP Langevin FINAL}).
The incoherent scattering events and the PL emission rates are
strongly dependent on the energy of the involved polariton states.
On the contrary parametric emission, being resonant when total
momentum is conserved, depends strongly on both the zenithal
$\theta$ and azimuthal $\phi$ angles which become poorly described
by such adaptive mesh when the dispersion curve becomes less steep.
Our DCTS-Langevin method enables the (computationally advantageous)
decoupling of incoherent dynamics from  parametric processes
allowing us to make the proper choices for the two contributions
whenever needed.

In particular, we seed the system at a specific ${\bf k}$ and first
of all we calculate the pump-induced PL by means of Eq.\,
(\ref{Pcroce P}). Because of the very steep dispersion curve and the
large portion of ${\bf k}$-space to be taken into account, in the
numerical solution we need to exploit the adaptive grid above
mentioned. Afterwards we use this pump-induced PL, $\langle
\hat{P}^\dag_{\bf k} \hat{P}_{\bf k} \rangle_{\text{PL}}$, as a
known input source in Eq.\, (\ref{PcrP Langevin FINAL}) where it is
largely more  useful to discretize uniformly in ${\bf k}$.

We consider a  SMC  analogous to that of Refs. \cite{Savasta
PRL2005,Langbein PRB2004} consisting of a $25$ nm $\text{Ga}
\text{As}/ \text{Al}_{0.3} \text{Ga}_{0.7} \text{As} $ single
quantum well placed in the center of a $\lambda$ cavity with
$\text{Al}\text{As}/$$\text{Al}_{0.15} \text{Ga}_{0.85} \text{As} $
Bragg reflectors. The lower polariton dispersion curve is shown in
Fig.\, 1. The simulations are performed at $T=5 \, \text{K}$ and the
measured cavity linewidth is $\hbar \gamma_c = 0.26\, \text{meV}$.
The laser pump is modeled as a single Gaussian-shaped impulse of
FWHM $\tau=1 \, \text{ps}$ exciting a definite wave vector ${\bf
k}_p$ and centered at $t=4\, \text{ps}$. We pump with co-circularly
polarized light exciting polaritons with the same polarization, the
laser intensities $I$ are chosen as multiple of $I_0$ corresponding
to a photon flux of $21 \, \mu\text{m}^{-2}$ per pulse. We observe
that Ref. \cite{Langbein PRB2004} excites with a linearly polarized
laser whose intensities $I^L$ are multiple of an $I^L_0$
corresponding to a photon flux of $21 \, \mu\text{m}^{-2}$ per pulse
too. In situations where the PSF and the MF terms dominate the
nonlinear parametric interaction, there  is no polarization mixing
and two independent parametric processes take place, the first
involving circularly polarized modes only and the second involving
counter-circularly polarized modes. Thus for comparison with theory
the effective density in those experiments is half the exciting
density: $I = I^L/2$.

It has been theoretically shown\cite{Piermarocchi bottleneck}, that
it is quite difficult to populate the polaritons in the strong
coupling region by means of phonon-scattering due to a bottleneck
effect, similar to that found in the bulk. Let us consider a pump
beam resonantly exciting polaritons at about the magic-angle.
Relaxation by one-phonon scattering events is effective when the
energy difference of the involved polaritons do not exceeds  $1\,
\text{meV}$. When polariton states within this energy window get
populated, they can relax by emitting a phonon to lower energy
levels or can emit radiatively. Owing to the reduced density of
states of polaritons and to the increasing of their photon-component
at lower energy, radiative emission largely exceeds phonon
scattering, hence inhibiting the occupation of the lowest polariton
states. Actually this effect is experimentally observed only very
partially and under particular circumstances\cite{Tartakovskii PRB}.
This is mainly due to other more effective scattering
mechanisms\cite{Di Carlo Kavokin PRB} usually present in SMCs. For
example the presence of free electrons in the system determines an
efficient relaxation mechanism. Here we present results obtained
including only phonon-scattering. Nevertheless the theoretical
framework here developed can be extended to include quite naturally
other enriching contributions that enhance non-radiative scattering
and specifically relaxation to polaritons at the lowest
k-vectors\cite{Di Carlo Kavokin PRB}. In order to avoid the
resulting  unrealistic low non-radiative scattering particularly
evident at low excitation densities, we artificially double the
acoustic-phonon scattering rates. However, acting this way, we
obtain non-radiative relaxation rates that in the mean agree with
experimental values.

We now present the results of numerical solutions of Eq.\,
(\ref{PcrP Langevin}) taking into account self-stimulation but
neglecting the less relevant pump-induced  renormalization of
polariton energies.

Figures\, 2 and 3 show the calculated time dependent polariton
mode-occupation of a signal-idler pair at  ${\bf k}=(0,0)$ and at
${\bf k}=(2\, k_m,0)$, respectively, obtained for four different
pump intensities in comparison with the time dependent pump-induced
PL at the corresponding ${\bf k}$. The pump beam is sent at the
magic angle \cite{Baumberg} ($k_m \simeq 1.44 \cdot 10^6\,
\text{m}^{-1}$) which is close to the inflection point of the energy
dispersion curve and is resonant with the polariton state at $k_m$.
The magic angle is defined as the pump value needed for the
eight-shaped curve of the resonant signal-idler pairs to intersect
the minimum of the polariton dispersion curve. It is worth noting
that the displayed results have no arbitrary units. We address
realistic input excitations and we obtain quantitative outputs,
indeed in Figs.\, 2 and 3  we show the calculated polariton
occupation, i.e. the number of  polaritons per mode. In our
calculations no fitting parameter is needed, nor exploited (apart
from the doubling of the phonon scattering rates). Moreover our
results predict in good agreement with the experimental results of
Ref.\, \cite{Langbein PRB2004} the pump intensity  at which
parametric scattering, superseding  the pump-PL, becomes visible. It
is clear the different pump-induced PL dynamics of the
mode-occupation at $k_x = 2\, k_m$ (Fig.\, 3) with respect to that
at the bottom of the dispersion curve of Fig.\, 2. Specifically a
residual queue at high time values, due to the very low radiative
decay of polaritons with k-vectors beyond the inflection point, can
be observed.  Furthermore we notice that already at moderate pump
excitation intensities the parametric contribution dominates. It
represents a clear evidence that we may device future practical
experiments exploiting such a window where the detrimental
pump-induced PL contribution is very low meanwhile we face a good
amount of polaritons per mode. Indeed for photon-counting
coincidence detections to become a good experimental mean of
investigation we need a situation where accidental detector's clicks
are fairly absent and where the probability of states with more than
one photon is low. Our results clearly show that there is a
practical experimental window where we would address a situation
where all these conditions would be well fulfilled.

We now focus our attention on the positive part of the $k_y=0$
section at different pump powers. In Fig.\, 4 we observe the clear
evidence of the build-up of the parametric emission taking over the
pump-induced PL once the seed beam has become enough intense, in
particular we can set a threshold around $I=10\,I_0$
($I^L=20\,I_0$). As expected, the parametric process with the pump
set at the magic angle enhances the specific signal-idler pair with
the signal in ${k_x=0}$ and the idler in ${k_x=2 k_m}$. We can
clearly see from the figure that at pump intensities higher than the
threshold the idler peak becomes more and more visible for
increasing power in agreement of what shown in Ref.\,\cite{Baumberg}
and Ref.\, \cite{Baumberg PRB(R)}. However, at so high $k_x$ values
the photon component is very small and even if the polariton idler
occupation is very high (as the inset if Fig.\, 4 shows), the
outgoing idler light is so weak to give some difficulties in real
experiments\cite{Langbein PRB2004}. Moreover we can notice that the
parametric process removes the phonon bottleneck in the region close
to ${\bf k}=0$. An analogous situation occurs also in Ref.\
\cite{Tartakovskii PRB}, though with a different SMC, where it can
be seen the bottleneck removal in ${\bf k}=0$ due to the parametric
emission.

Fig.\, 5  shows the impact on the time integrated patterns of the
calculated pump-induced PL. We consider, for different excitation
intensities, the solutions of Eq.\, (\ref{PcrP Langevin FINAL}) with
and without the pump-induced PL occupations. As can be seen its
inclusion does not result in an uniform noise background, but it
seems to somewhat remember its incoherent nonuniform distribution
(the one depicted in the corresponding curve, i.e.
$\text{PL}_{\text{pump}}$, in Fig.\, 4). As can be clearly gathered
from the figure, the pump-induced PL has a non negligible
contribution in a region in ${\bf k}$-space resonant for the
parametric processes. As a consequence at intermediate excitation
intensities it adds up to the parametric part reaching a
contribution even comparable to the peak of emission set in ${\bf
k}=0^+$. Only beyond the above mentioned threshold the parametric
emission is able to take over pump-induced PL and results in the
great emission in the bottom of the dispersion curve of Ref.\
\cite{Baumberg}. These results clearly show that PL emission does
not become negligible at quite high excitation densities, but, being
amplified by the parametric process, determines a redistribution of
polariton emission displaying qualitative differences with respect
to calculations neglecting PL. An interesting question regarding
these phenomena could be related to the impact in the global
spontaneous emission of the two contributions in Eq.\, (\ref{PcrP
Langevin FINAL}), namely that of the homogeneous part $|c_2(0,t)|^2$
and the one originating from noise operators in the time integral in
the last line. In the inset of Fig.\, 5 we have depicted the ratio
of the homogenous solution with the global emission at ${\bf
k}=0^+$, calculated without $\langle \hat{P}^\dag_{\bf k}
\hat{P}_{\bf k} \rangle_{\text{PL}}$. Rather surprisingly when
increasing the pump intensity the two contributions (homogeneous and
particular) continue to have comparable weights, hence for a proper
description of the spontaneous parametric emission they must be both
included.

The calculated time-integrated spectra of the outgoing light at
${\bf k}=(0,0)$ obtained at six different pump intensities for an
excitation at the magic angle $k_m$ are shown in Fig.\, 4. It can be
easily noticed a threshold around $I^L=20\,I_0$ in perfect agreement
with the results in Fig.\, 4 and with Ref.\, \cite{Langbein
PRB2004}. For intensities lower than the threshold, the signal in
${k_x=0}$ (with the corresponding idler in ${k_x=2 k_m}$) shows a
quite large nearly Lorentian shape. As soon as the threshold is
passed over, the spectrum starts to increase super-linearly with
some spurious queues due to (calculated) asymmetric signal/idler
damping values. Noticeably the spectra show an evident linewidth
narrowing for increasing pump intensities witnessing the parametric
emission build-up. For the sake of presentation in the inset we also
present some normalized spectra which give immediate evidence of the
build-up of a narrow linewidth beyond the mentioned threshold.

\section{Summary and conclusions}\label{VII}
Based on a DCTS theoretical framework for interacting polaritons
(see Sect. 2), we have presented a  general theoretical approach for
the realistic investigation of polariton quantum correlations in the
presence of coherent and incoherent interaction processes. The
proposed theoretical framework combines the dynamics controlled
truncation scheme with the nonequilibrium quantum Langevin approach
to open systems. It provides an easy recipe to calculate multi-time
correlation functions which are key-quantities in quantum optics,
but as shown here even for single-time quantities it provides a
natural and advantageous decoupling of incoherent dynamics from
parametric processes. We have elaborated equations whose structure
is analogous to those one obtains by means of bosonization
\cite{Ciuti SST}. However, thanks to the DCTS approach we have been
able to obtain microscopically nonlinear coefficients with great
accuracy. In particular in Ref. \cite{Ciuti SST} the nonlinear
coupling coefficient contains additional terms due to phase space
filling providing an interaction strength larger of about a factor
3. We believe that the difference is due to the way how the
Bosonization procedure has been adopted. As a first application of
the proposed theoretical scheme, we have analyzed the build-up of
polariton parametric emission in semiconductor microcavities
including the influence of noise originating from phonon induced
scattering. Our numerical results clearly show the importance of a
proper microscopic analysis able to account for parametric emission
and pump-induced PL on an equal footing in order to make
quantitative comparison and propose future experiments, seeking and
limiting all the unwanted detrimental contributions. Specifically,
we have shown that already at moderate pump excitation intensities
there are clear evidence that we may device future practical
experiments exploiting existing situations where the detrimental
pump-induced PL contribution is very low meanwhile we face a good
amount of polaritons per mode. It represents an exciting and
promising possibility for future coincidence experiments even in
photon-counting regimes, vital for investigating nonclassical
properties of the emitted light.

\appendix*
\section{The interactions with reservoirs}\label{exc2pol}
A quasi-two-dimensional exciton state with total in-plane center of
mass (CM) wave vector ${\bf k}$ may be represented as
\cite{Takagahara}
\begin{equation}\label{exciton W}
\ket{\lambda, {\bf k}_{}}\!\! =\!\! \frac{v_0}{\sqrt{S}}\!\!
\sum_{\bf r_e,r_h} \!\! e^{i \bf k_{} \cdot R^{\|}} F_{\lambda}({\bf
r^{\|}_{e} - \bf r^{\|}_{h}},z_e,z_h) a^\dag _{c,{\bf r_e}}a_{v,{\bf
r_h }}\!\! \ket{0}\, ,\end{equation} where $v_0$ and $S$ are the
volume of the unit cell and the in-plane quantization surface,
whereas $a^\dag _{c/v,{\bf r}} (a_{c/v,{\bf r }})$ are creation
(annihilation) operator of the conduction- or valence- band electron
in the Wannier representation. ${\bf r}_{e/h} = ({\bf
r}^{\|}_{e/h},z_{e/h})$ are to be considered coordinates of the
direct lattice, $\ket{\!\!0}$ is the crystal ground state and ${\bf
R}$ the exciton center of mass coordinate ${\bf R} = (m_e {\bf r}_e
+ m_h {\bf r}_h)/(m_e + m_h)$ with $m_e$ and $m_h$ the effective
electron and hole masses. The Wannier envelope function
$F_{\lambda}$ is normalized so that the integral over the whole
quantization volume $(V \!\! = S\!\! \cdot \!\! L)$ of its square
modulus is equal to 1. The lattice properties of GaAs-AlAs QW
structures are in close proximity, thus the acoustic-phonons which
interact with the quasi-two-dimensional exciton can be considered to
have three-dimensional character. The electron-phonon interaction
Hamiltonian resulting  from the deformation potential coupling, can
be written as
\begin{eqnarray}\label{H eh-ph} H^{DF}_{e-ph} = \sum_{\bf k,q} && \left( \frac{\hbar
|{\bf q}|}{2 \rho u V} \right)^{1/2} \left( D_c a^\dag _{c,{\bf k +
q}}a_{c,{\bf k }} + D_v a^\dag _{v,{\bf k + q}}a_{v,{\bf k }}
\right) \left( b_{\bf q} + b^\dag_{- \bf q} \right)\,
.\end{eqnarray} Here $a^\dag _{c/v,{\bf k}}, a_{c/v,{\bf k }}$ are
creation and destruction operator of the conduction- valence- band
electron in Bloch representation. Transforming from the Wannier to
the Bloch representation, we shall project (\ref{H eh-ph}) into the
excitonic bases. Moreover, since we are interested in the 1S exciton
sector $\lambda = (n, \sigma)$ only
\begin{eqnarray} &&F_{1 S} = W_{1
S}({\bf r^{\|}_{e} - \bf r^{\|}_{h}})\chi_{e}(z_e)\chi_{h}(z_h) \nonumber \\
&&W_{1 S}(\rho) = \sqrt{2/(\pi a^2_\text{x})}
\exp(-\rho/a^2_\text{x})\, .\end{eqnarray} It yields
\begin{eqnarray}\label{H exc-ph}
H^{DF}_{exc-ph} =&& \hspace{-0.3cm} \sum_{{\bf k, k'},q_z}
\hspace{-0.0cm} t^{{\bf k}',q_z}_{{\bf k}}\ket{1 S\,{\bf k'}} \bra{1
S\,{\bf k}} \left( b_{({\bf k' - k},q_z)} + b^\dag_{-({\bf
k'-k},q_z)} \right)\, , \end{eqnarray} here \begin{eqnarray}\label{t
coeff} t^{{\bf k}'\!,q_z}_{{\bf k}}\!\!\!&=& \!\! \left( \frac{\hbar
\sqrt{|{\bf k' - k}| + q^2_z}}{2 \rho u V} \right)^{1/2}
\hspace{-0.5cm} \left( D_c I^\bot_e(q_z) I^\|_e({\bf k'-k}) -  D_v
I^\bot_h(q_z) I^\|_h({\bf k'-k}) \right)\, ,
\end{eqnarray}
being $I^\bot$ and $I^\|$ overlap integrals given in Eq.\,
(\ref{over}).

We  treat the cavity field in the quasi-mode approximation, that is
to say we shall quantize the field as the mirror were perfect and
subsequently we shall couple the cavity with a reservoir of a
continuum of external modes. The coupling of the electron system and
the cavity modes is given in the usual rotating wave approximation
\begin{equation}\label{Ham inter cav-qm} H_{qm} \!\!= i \hbar \sum_{\bf k}
\!\!\int \!\!\!d\omega g_{\bf k}(\omega) a^{\dag}_{\bf k}
E^{(-)}_{\bf k}(\omega,t)+ H.c.\, ,\end{equation}
In passing form the air to the SMC, we change from a 3D to a 2D
quantization, it means that in the coupling once either $({\bf
k},k_z)$ or $({\bf k},\omega)$ is chosen the third follows
consistently. We have chosen the latter for simplicity in dealing
with the Markov machinery. In the Hamiltonian $g_{\bf k}(\omega)$ is
the coupling coefficient, a sort of \textit{optical matrix element},
$E^{(-)}_{\bf k}(\omega,t)$ and $E^{(+)}_{\bf k}(\omega,t)$ are the
two propagating normal modes of the external light. Modeling the
loss through the cavity mirrors within the quasi-mode picture means
we are dealing with an ensemble of external modes, generally without
a particular phase relation among themselves. An input light beam
impinging on one of the two cavity mirrors is an external field as
well and it must belong to the family of modes of the corresponding
side (i.e. left or right). It will be nothing but the non zero
expectation value of the (coherent) photon operator giving a non
zero contribution on the $1^{\text{st}}$ perturbative order. All the
other incoherent bath modes will have their proper contribution in
the $2^{\text{nd}}$ order calculations.

It is worth noting that the treatment of the cavity losses as a
scattering interaction is a result of the form chosen of the
effective quasi-mode Hamiltonian. However, even if a model
Hamiltonian, the quasi-mode description has given a lot of evidence
as an accurate modeling tool and it is widely used in the
literature. Let us call $R$ the quasi-mode reservoir Hamiltonian. It
can be shown that the first order (coherent) dynamics for a generic
operator $\hat{O}$ under the influence of the coherent part of the
quasi-mode ensemble reads, (see Eq.\, (\ref{Ham inter cav-qm}))

\begin{eqnarray}\label{qm 1rst} &&i \hbar \frac{d \langle \hat O\rangle}{dt}{\left|\right._{H_{qm}}} =i \hbar \sum_{{\bf k}} \sum_{p} g_{p} \left<\right.
E^{(-)}_p(\Omega_p,t) \left.\right>_R  \left<\right.
[\hat{O},a^\dag_{\bf k} ] \left.\right> + H.c.\, ,
\end{eqnarray} where $g_{\bf k}(\omega)$ is the coupling
coefficient, $E^{(-)}_{\bf k}(\omega,t)$ is the propagating normal
mode of the external light and $\sum_{p} g_{p} \left<\right.
E^{(-)}_p(\Omega_p,t) \left.\right>_R$ is the superposition of all
the possible coherent pump feeds.

An interesting situation occurs within the assumption of a flat
quasi-mode spectrum, an approximation almost universally made in
quantum optics \cite{Collet Gardiner PRA 1985}. It makes Eq.\,
(\ref{gamma rad}) independent of the frequency:

\begin{equation} \gamma^{(c)}_{\alpha,{\bf k}} = \sum_{i=l,r}
\frac{2 \pi}{\hbar} |C_{\alpha,{\bf k}}|^2 \hbar |g_{i,{\bf k}}|^2 =
|C_{\alpha,{\bf k}}|^2 \sum_{i=l,r} \gamma^{(m)}_{i,{\bf k}}\,
,\end{equation} where $\gamma^{(m)}_{i,{\bf k}}$ is the (i-side)
damping of the cavity without the quantum well.

Thus \begin{equation}\label{def tc} \sum_{i=l,r}
\gamma^{(m)}_{i,{\bf k}} = 2 \pi \sum_{i=l,r}|g_{i,{\bf k}}|^2\,
,\end{equation} there are two situations:
\begin{itemize}
  \item equal damping: $\gamma^{(m)}_{{l,\bf k}} = \gamma^{(m)}_{{r,\bf
  k}}$, we can define the transmission coefficient of the i-side
  \begin{equation} |g_{i,{\bf k}}|^2 = \frac{\gamma^{(m)}_{{i,\bf
  k}}}{2 \pi} \doteq t^2_{c,i} \end{equation}
  \item we know the ratio:
  \begin{displaymath}
\left\{ \begin{array}{ll}
R=\frac{\gamma^{(m)}_{{r,\bf k}}}{\gamma^{(m)}_{{l,\bf k}}} \\
\gamma^{(m)}_{{\text{tot},\bf k}} = \gamma^{(m)}_{{r,\bf k}} + \gamma^{(m)}_{{l,\bf k}}\\
\end{array} \right. \Rightarrow \left\{ \begin{array}{ll}
\gamma^{(m)}_{{l,\bf k}} = \frac{1}{1+R} \gamma^{(m)}_{{\text{tot},\bf k}} \\
\gamma^{(m)}_{{r,\bf k}} = \frac{R}{1+R} \gamma^{(m)}_{{\text{tot},\bf k}}\\
\end{array} \right.
\end{displaymath}
\end{itemize}and the transmission coefficients follow.

In the light of the definition of $t^2_{c,i}$, it becomes evident
that the semiclassical coherent input feed could also be modeled
from the beginning with an effective Hamiltonian:
\begin{equation} H_p = i \hbar \sum_{{\bf k}}
({E}^{(-)}_{{\bf k}} \hat{a}^\dag_{{\bf k}} - {E}^{(+)}_{{\bf k}}
\hat{a}_{{\bf k}} )\, ,
\end{equation} where (the $\mathbb{C}-$numbers) ${E}^{(\pm)}_{{\bf k}} = \sum_{p}
t_{c,p} \left<\right. E^{(\pm)}_p(\Omega_p,t) \left.\right>_R$
represent the incoming coherent input beams \cite{Savasta PRL96}.

\newpage

\begin{figure}[!ht]
\begin{center}
\resizebox{!}{!}{
\includegraphics{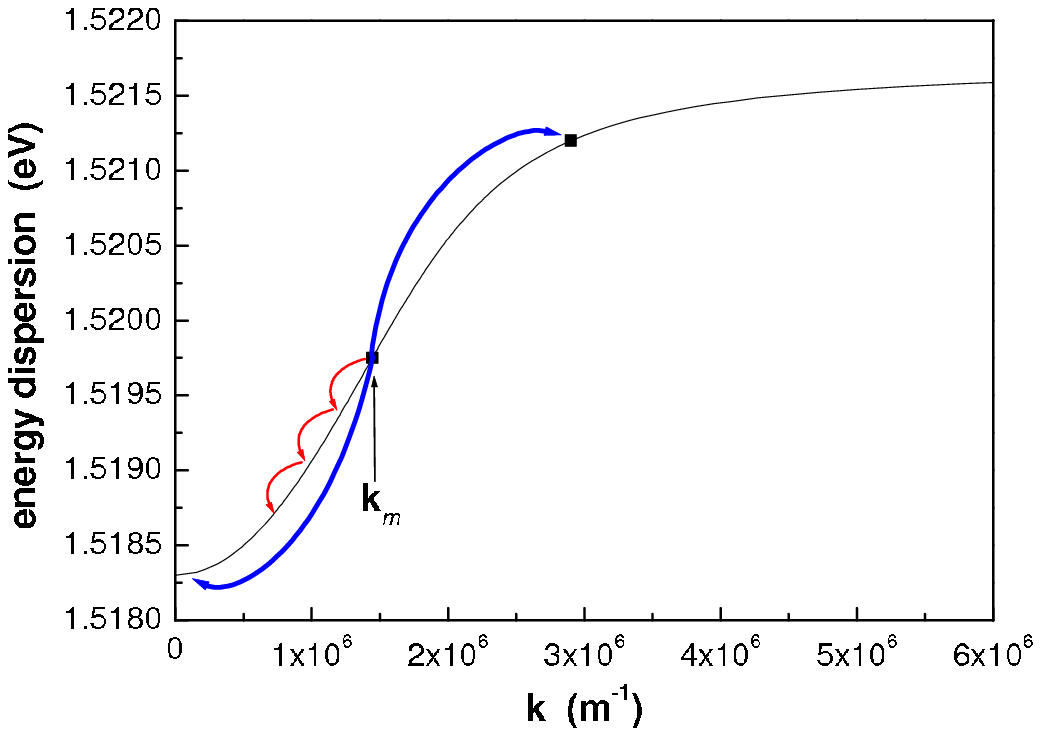}}\caption{(color online) Energy dispersion of the lowest polariton branch for
the structure of Ref.\, \cite{Langbein PRB2004} consisting of a $25$
nm $\text{Ga} \text{As}/ \text{Al}_{0.3} \text{Ga}_{0.7} \text{As} $
single quantum well placed in the center of a $\lambda$ cavity with
$\text{Al}\text{As}/$$\text{Al}_{0.15} \text{Ga}_{0.85} \text{As} $
Bragg reflectors. The pump at the magic angle and its parametric
scattering (blue curve) are schematically depicted. The latter
scatters two pump polaritons in a signal-idler couple at ${\bf k}=0$
and ${\bf k} = 2{\bf k}_m$. The red curve symbolizes incoherent pump
scattering at $\text{T} = 0\, \text{K}$, e.g. due to acoustic phonon
interaction. Because of the very steep dispersion curve in the
strong coupling region due to strong coupling, for a pump beam set
at the magic angle or beyond, even a small temperature of $5$ K is
sufficient to enable scattering processes towards states at quite
higher $k$-vectors, thus it becomes necessary to include a
computational window in k-space, significantly beyond ${\bf
k}_{pump}$ making difficult numerical simulations.}\label{disp}
\end{center}
\end{figure}

\newpage

\begin{figure}[!ht]
\begin{center}
\resizebox{!}{9.0cm}{
\includegraphics{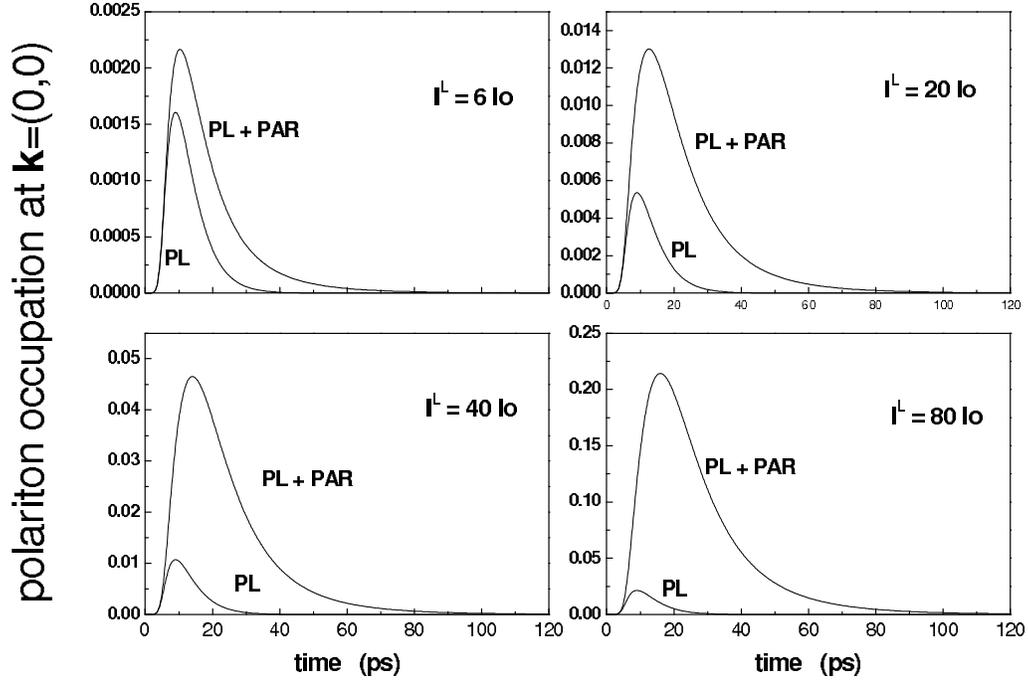}}\caption{Calculated time dependent polariton
mode-occupation at ${\bf k}=(0,0)$ obtained at four different pump
intensities  in comparison with the time dependent pump-induced PL
at the corresponding ${\bf k}$. The laser pump is modeled as a
single Gaussian-shaped impulse of FWHM $\tau=1 \, \text{ps}$
exciting a definite wave vector ${\bf k}_p= (k_m,0)$ and centered at
$t=4\, \text{ps}$. The calculated values are in  good quantitative
agreement with the measured value of Ref.\cite{Langbein PRB2004},
with no fitting parameter needed nor exploited.}\label{NkS di t}
\end{center}
\end{figure}

\bigskip

\begin{figure}[!ht]
\begin{center}
\resizebox{!}{9.0cm}{
\includegraphics{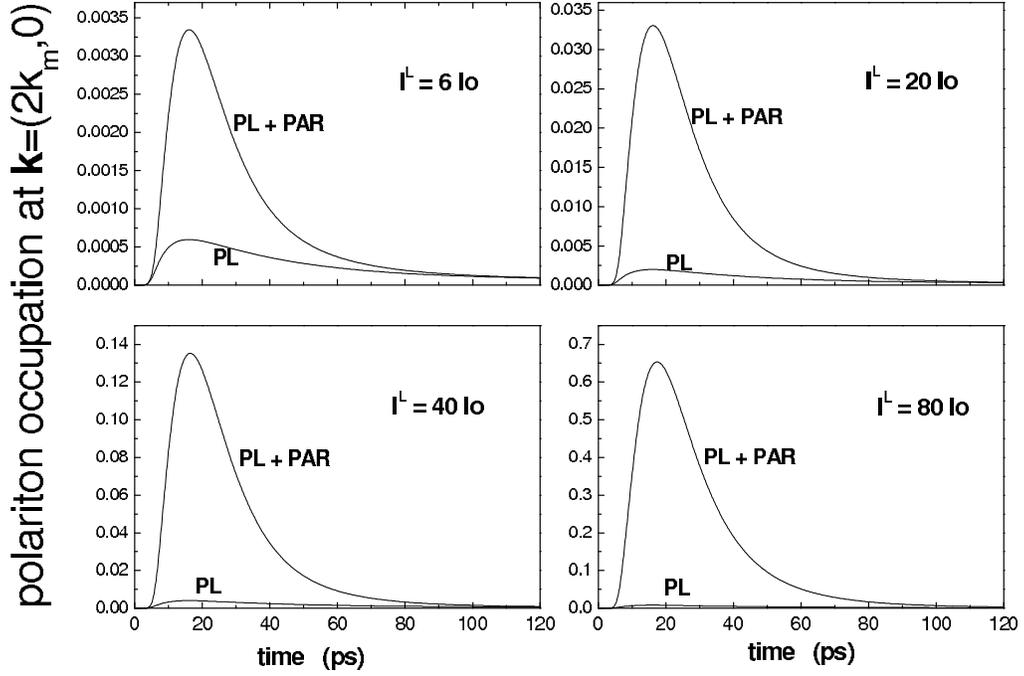}}\caption{Calculated time dependent polariton
mode-occupation at ${\bf k}=(2\,k_m,0)$. It has been obtained under
the same condition as Fig.\, 2. It is clear the different
pump-induced PL dynamics with respect to Fig.\, \ref{NkS di t},
specifically a residual queue at high time values.}\label{NkI di t}
\end{center}
\end{figure}

\newpage

\begin{figure}[!ht]
\begin{center}
\resizebox{!}{9.0cm}{
\includegraphics{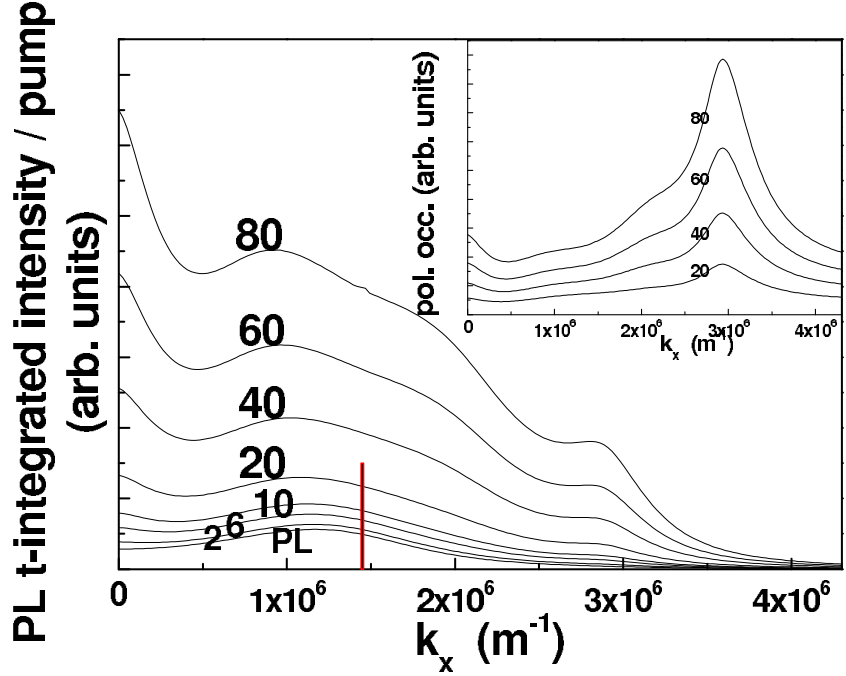}}\caption{ Time-integrated outgoing photon emission
intensity. The pump is set at ${\bf k}_p = (k_m,0)$. It is clear the
evidence of the build-up of the parametric emission taking over the
pump-induced PL once the seed beam becomes higher than the threshold
around $I^L=20\,I_0$. Moreover the parametric process removes the
phonon bottleneck in the region closed to ${\bf k}=0$. As expected
the specific signal-idler parametric scattering with the signal in
${k_x=0}$ and the idler in ${k_x=2 k_m}$ is favourite and at higher
pump intensities dominates the light emission. The polariton idler
occupations for some pump values are depicted in the inset. Although
polariton occupation at ${k_x=2 k_m}$ is so high, its photonic
component is very small resulting in a very weak outgoing light
beam.}\label{tintegrated ky=0}
\end{center}
\end{figure}

\newpage

\begin{figure}[!ht]
\begin{center}
\includegraphics{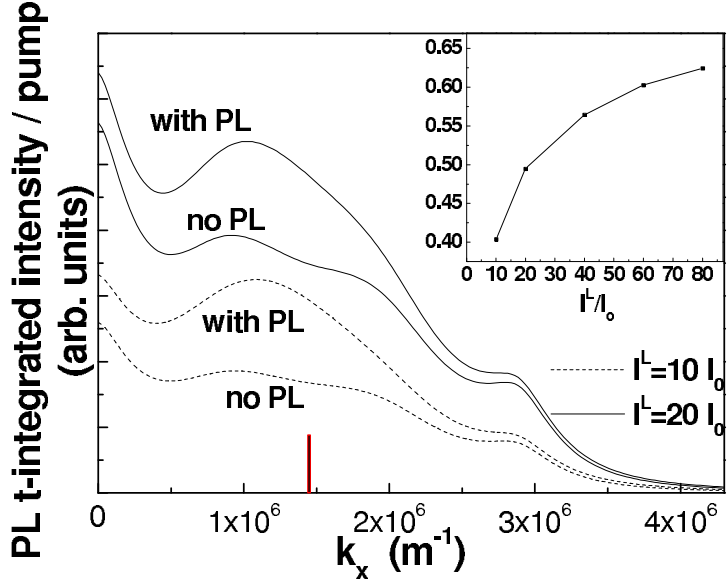}\caption{ The impact on the time integrated
patterns of the calculated pump-induced PL for different excitation
intensities is shown (the pump is set in ${\bf k}_p = (k_m,0)$).
Here the solutions of Eq.\, (\ref{PcrP Langevin FINAL}) with and
without the pump-induced PL occupations are depicted. On the
contrary to what implicitly considered in previous phenomenological
theories it results in a non-uniform noise background and hence its
momentum distribution has to be included for a realistic microscopic
calculation of the emission patterns. Moreover it is non-negligible
in a region in ${\bf k}$-space resonant for the parametric processes
and hence at intermediate excitation intensities it adds up to the
parametric part reaching a contribution even comparable to the peak
of emission set in ${\bf k}=0^+$ up to the threshold around
$I^L=20\, I_0$. In the inset the ratio of the homogenous solution
with the global emission at ${\bf k}=0^+$ (both calculated without
$\langle \hat{P}^\dag_{\bf k} \hat{P}_{\bf k} \rangle_{\text{PL}}$)
is depicted. For increasing pump intensities the two contributions
(homogeneous and particular) in Eq.\, (\ref{PcrP Langevin FINAL})
still display comparable contributions, hence for a proper
description of the spontaneous parametric emission they must be both
included.}\label{parVSpl}
\end{center}
\end{figure}

\newpage

\begin{figure}[!ht]
\begin{center}
\resizebox{!}{9.0cm}{
\includegraphics{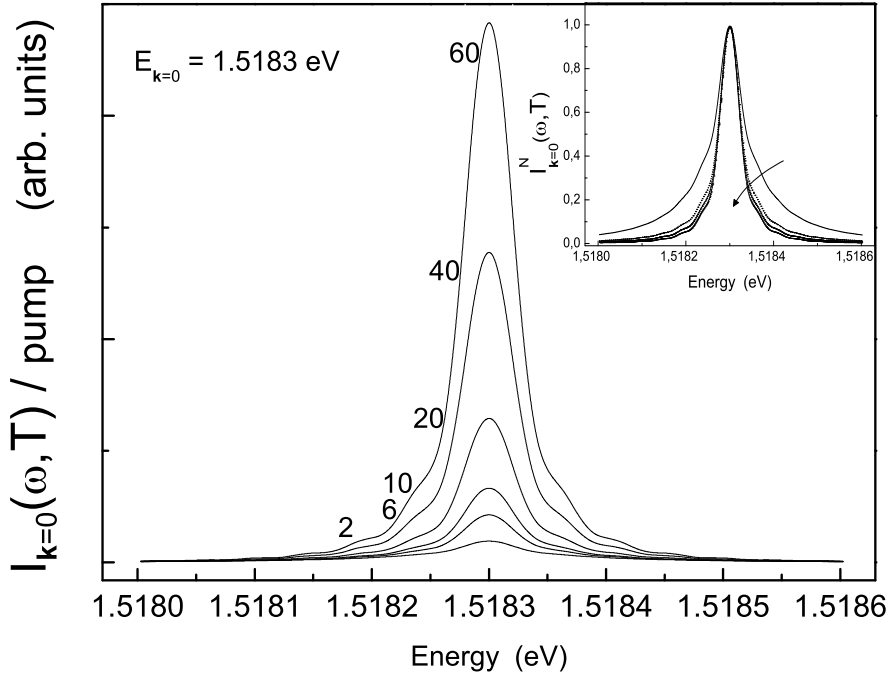}}\caption{Calculated time-integrated spectrum
of the outgoing light at ${\bf k}=(0,0)$ normalized with respect to
the pump seed obtained at six different pump intensities for an
excitation at the magic angle $k_m$. It can be easily noticed a
threshold around $I^L=20\,I_0$ in perfect agreement with the results
in Fig.\, 4 and with Ref.\, \cite{Langbein PRB2004}. For intensities
lower than the threshold, the signal in ${k_x=0}$ (with the
corresponding idler in ${k_x=2 k_m}$) shows a quite large nearly
Lorentian shape, as soon as the threshold is passed over the
spectrum  start to increase super-linearly meanwhile the linewidth
decreases witnessing the parametric emission build-up. In the inset
the normalized spectra at increasing pump powers (indicated by the
arrow direction) are depicted evidencing even better the linewidth
narrowing. We notice also some spurious queues due to (calculated)
asymmetric signal/idler damping values.}\label{spettro}
\end{center}
\end{figure}

\end{document}